\documentclass[12pt,preprint2]{aastex}

\usepackage[T1]{fontenc}
\usepackage{lmodern}
\usepackage{breqn}
\usepackage{graphicx}
\usepackage{txfonts}
\usepackage{lscape}
\usepackage{aas_macros}
\usepackage{amssymb}
\usepackage{natbib}
\bibpunct{(}{)}{;}{a}{,}{,}
\usepackage{longtable}
\usepackage[hang,bf,footnotesize]{subfigure}
\usepackage[figuresright]{rotating}
\begin{document}

\title{Photoevaporation and close encounters: how the environment around Cygnus~OB2 affects the evolution of protoplanetary disks}

\author{M. G. Guarcello\altaffilmark{1,2}, J. J. Drake\altaffilmark{2}, N. J. Wright\altaffilmark{3,2}, J. F. Albacete-Colombo\altaffilmark{10}, C. Clarke\altaffilmark{11}, B. Ercolano\altaffilmark{5,6}, E. Flaccomio\altaffilmark{1}, V. Kashyap\altaffilmark{2}, G. Micela\altaffilmark{1}, T. Naylor\altaffilmark{7}, N. Schneider\altaffilmark{8}, S. Sciortino\altaffilmark{1}, J. S. Vink\altaffilmark{9}}

\altaffiltext{1}{INAF - Osservatorio Astronomico di Palermo, Piazza del Parlamento 1, I-90134, Palermo, Italy\\
              \email{mguarce@astropa.unipa.it}}
\altaffiltext{2}{Smithsonian Astrophysical Observatory, MS-67, 60 Garden Street, Cambridge, MA 02138, USA}
\altaffiltext{3}{CAR/STRI, University of Hertfordshire, College Lane, Hatfield, AL10 9AB, UK}
\altaffiltext{4}{School of Physics \& Astronomy, Cardiff University, The Parade, Cardiff CF24 3AA, UK}
\altaffiltext{5}{Universit\"{a}ts-Sternwarte M\"{u}nchen, Scheinerstrasse 1, D-81679 M\"{u}nchen, Germany}
\altaffiltext{6}{Excellence Cluster Universe, Boltzmannstr. 2, D-85748 Garching, Germany}
\altaffiltext{7}{School of Physics, University of Exeter, Stocker Road, Exeter EX4 4QL, UK}
\altaffiltext{8}{Physik Institut, University of Cologne, 50937 Cologne, Germany}
\altaffiltext{9}{Armagh Observatory, College Hill, BT61 9DG Armagh, UK}
\altaffiltext{10}{Sede Atlantica de la Universidad Nacional de Rio Negro, Don Bosco y Leloir s/n, 8500 Viedma RN, A}
\altaffiltext{11}{Institute of Astronomy, Madingley Rd., Cambridge, CB3 0HA, UK}
\begin{abstract}

In our Galaxy, star formation occurs in a variety of environments, with a large fraction of stars formed in clusters hosting massive stars. OB stars have an important feedback on the evolution of protoplanetary disks orbiting around nearby young stars and likely on the process of planet formation occurring in them. The nearby massive association Cygnus~OB2 is an outstanding laboratory to study this feedback. It is the closest massive association to our Sun, and hosts hundreds of massive stars and thousands of low mass members, both with and without disks. In this paper, we analyze the spatial variation of the disk fraction (i.e. the fraction of cluster members bearing a disk) in Cygnus~OB2 and we study its correlation with the local values of Far and Extreme ultraviolet radiation fields and the local stellar surface density. We present definitive evidence that disks are more rapidly dissipated in the regions of the association characterized by intense local UV field and large stellar density. In particular, the FUV radiation dominates disks dissipation timescales in the proximity (i.e. within $0.5\,$pc) of the O stars. In the rest of the association, EUV photons potentially induce a significant mass loss from the irradiated disks across the entire association, but the efficiency of this process is reduced at increasing distances from the massive stars due to absorption by the intervening intracluster material. We find that disk dissipation due to close stellar encounters is negligible in Cygnus~OB2, and likely to have affected 1\% or fewer of the stellar population. Disk dissipation is instead dominated by photoevaporation. We also compare our results to what has been found in other young clusters with different massive populations, concluding that massive associations like Cygnus~OB2 are potentially hostile to protoplanetary disks, but that the environments where disks can safely evolve in planetary systems are likely quite common in our Galaxy. 

\end{abstract}


\section{Introduction}
\label{intro}

The evolution of circumstellar disks around young stars is symbiotic with the formation of planetary systems. In the generally accepted paradigm of star formation, about 80\% of the stars younger than $1\,$Myr are surrounded by a thick circumstellar disk, composed of dust and gas in a typical mass ratio of 1:100 and emitting in the near-infrared. This fraction decreases to $\sim55\%$ at $3\,$Myrs and $\sim15\%$ at $\sim5\,$Myrs \citep{HaischLL2001, HernandezHMG2007, Mamajek2009}. Disks are generally dissipated, then, in less than $10\,$Myrs, although this timescale has been brought into question by recent studies \citep{BellNMJ2013, DeMarchiPGB2013}. This timescale is long enough to allow the early steps of planet formation to occur, such as settling of dust grains on the disk midplane, aggregation of dust grains into larger solid bodies and the formation of planetesimals (i.e. \citealp{Zuckerman2001, DullemondDominik2004}).{\bf Consequently, stars that have dissipated their disks quickly (i.e. in less than 2 or 3 Myrs) may be less likely to form planets than disks which evolved unperturbed}. \par

  The evolutionary timescale of circumstellar disks is dictated by several processes. Viscosity in the disk drives a radial flow of gas that finally accretes onto the central star (see, e.g., \citealp{Pringle1981}). In the very inner part of the disk, the accretion is controlled by the stellar magnetic field, which funnels the plasma in accretion columns \citep{GullbringHBC1998,RomanovaLLK2011}. The accreting material hits the stellar surface almost in free-fall at speeds of a few hundred km/s, heating the accretion spots up to temperatures of approximately ten thousand degrees \citep{CalvetGullbring1998}. This process is a source of energetic ultraviolet and soft X-ray radiation. This energetic radiation (together with the intense coronal X-ray emission which characterizes young stars; \citealp{Montmerle1996}) has an important impact on the evolution of the circumstellar disk itself. Far ultraviolet ($FUV$) photons (with energy between $6\,$eV$\,<\,$h$\nu\,<13.6\,$eV) dissociate H$_2$ molecules, while the extreme ultraviolet ($EUV$) and X-ray photons (with energy h$\nu\,>13.6\,$eV) are capable of ionizing hydrogen atoms. This input of energy raises the gas temperature in the outer layers in the disk up to some thousand degrees, and sometime even more than ten thousand, creating an intense thermal pressure that drives a flow of gas away from the disk. This process, called ``{\it disk photoevaporation}'' (e.g. \citealp{JohnstoneHB1998}), is thought to be responsible for the clearing of an intermediate region in the disk in few Myrs and the formation of pre-transition disks \citep{ClarkeGS2001,AlexanderCP2006,ErcolanoDRC2008}. Once the inner disk is decoupled from the outer disk, the inner part accretes onto the central star in a viscous timescale of few $10^5\,$yrs, leading to the creation of transition disks \citep{CalvetDWF2005}. \par

Disk photoevaporation is then a crucial process in normal disk evolution, occurring in less than $10\,$Myrs, when it is induced by the central star itself, and it may also impact planets formation and migration across the disk. However, a significant fraction of stars in our Galaxy form in the proximity of OB stars, which are intense sources of UV radiation and may induce photoevaporation in nearby disks. For instance, in the Trapezium Cluster in Orion disk photoevaporation externally induced by the O6V star $\Theta^1$ Ori has been invoked to explain the protoplanetary disks embedded in an evaporating envelope observed by the Hubble Space Telescope \citep{ODellW1994}. Similar structures have been observed in other clusters, such as the proplyd-like objects in Cygnus~OB2 \citep{WrightDDG2012}, some of which were shown to be evaporating disks by \citet{GuarcelloDWG2014}. In the intermediately massive clusters (i.e. hosting tens of massive stars) NGC~6611 \citep{GuarcelloPMD2007, GuarcelloMDP2009, GuarcelloMPP2010} and NGC~2244 \citep{BalogMRS2007}, the fraction of members with disks has been observed to decline in the proximity of the massive cluster members, as a consequence of a fast disk dissipation due to the intense local UV field. Such externally induced photoevaporation can dissipate disks in timescales as short as $\sim1\,$Myr \citep[i.e.][]{StorzerHollenbach1999}. In low mass clusters hosting only a few massive stars, disks are expected to be affected by the UV radiation field only in proximity (i.e. $\ll 1 \,$pc) of OB stars.  \par
  
  Externally induced photoevaporation is not the only environmental feedback that can affect disk evolution. During the dynamical evolution of clusters stars can occasionally encounter other members at small distances. In those cases when the impact parameter is smaller than a few hundred AU and one or both of the interacting stars has a disk, the close encounter can have dramatic effects on the morphology and evolution of the disk, resulting in: significant mass loss from the disk, with part of the material being dispersed in the surrounding medium and part captured by the other star \citep{ClarkePringle1993,PfalznerUH2005,ThiesKGS2010}; enrichment of one circumstellar environment as a result of the mass exchange with more evolved systems \citep{AdamsS2005}; expulsion of forming planets from the disk \citep{AdamsLaughlin2001,SpurzemGHL2009}, that may result in floating planets, or perturbation of the orbits of forming planetesimals and planets \citep[i.e. ][]{ZapateroOsorioBMR2000}. The mass loss from the disk during a close encounter depends on several factors: masses and velocities of the interacting stars, the impact parameter, the direction of motion of the interacting stars with respect to the orbital motion in the disk, the angle between the plane of the disk and the direction of the stellar interaction, etc... \par
  
Several theoretical studies have examined the impact of close encounters on the disk population in clusters with different stellar density. For instance, \citet{AdamsPFM2006} found that in small clusters with a population of a few hundred members, stars have a probability between 0.1\% to 1\%  of having a close encounter with an impact parameter of $100-200\,$AU in one Myr, and concluded that in low-mass environments disks are rarely dispersed by the gravitational interaction between stars. \citet{PfalznerOE2006} have simulated the gravitational interactions in intermediately massive environments such as the Trapezium in Orion. They found that, on average, in $10\,$Myrs disks around massive stars ($M_{stars}>10\,$M$_{\odot}$) lose $\sim80\%$ of their initial mass, disk in stars with $1\,M_{\odot}<M_{stars}<10\,$M$_{\odot}$ about 30\%, while less massive stars lose about 20\%. In this environment, the chances for very close encounters (i.e. $b<90\,$AU) for a solar mass star has been calculated by \citet{Adams2010} to range between 1\%-10\% in $10\,$Myrs. The hybrid N-body/SPH (smoothed particle hydrodynamic) simulation presented in \citet{RosottiDDH2014} have shown that close encounters in clusters mainly affect disks size than disks mass. In the most extreme environments, such as the Arches cluster where the central 1$\,$pc cubic region contains about 125 O stars, one third of the disks are destroyed by gravitational interactions in less than $3\,$Myrs \citep{OlczakKHP2012}. There is a wide range of situations, then, suggested by these theoretical studies, but which still lack observational support. \par
  
   As we discuss in Section \ref{cygob2_sect}, Cyg~OB2 is dynamically not evolved, rich of O stars and low-mass stars both with and without disks, thus particularly well-suited to the study of disk photoevaporation, also because the radiation exposure of the low mass stars can be estimated reliably. This is not the case in most of the known massive clusters, which are more distant, more dynamically evolved and mixed. In this paper we study the evolutionary timescale for disk dissipation in Cygnus~OB2, and we compare our results to previous studies focused on the same topic but in different star forming regions. In Sect \ref{cygob2_sect} we describe Cygnus~OB2 and the data from the Cygnus~OB2 Chandra Legacy Project; the evidence that disks are dissipated by the environment feedback in Cyg~OB2 are presented in Sect. \ref{df_sect}, and in Sect \ref{map_sec} we compare the effectiveness of externally induced photoevaporation and stellar close encounters on disk dissipation. Finally, in Sect. \ref{disc_sec} we discuss the importance of the environment feedback on disk evolution and planet formation, first in Cygnus~OB2 and then in other star-forming environments of our Galaxy.\par

\section{Cygnus~OB2 and the Cyg~OB2 Chandra Legacy Survey}
\label{cygob2_sect}

The massive association Cygnus~OB2 in the Cygnus~X complex provides a laboratory to test the effects of environmental feedback on the evolution of circumstellar disks and planet formation. At a distance of $1400\,$pc \citep{RyglBSM2012} it is the closest massive young association to the Sun. The massive population of Cyg~OB2 has been the subject of several studies  \citep{ReddishLP1967, Knodlseder2000, ComeronPRS2002, Hanson2003, DrewGIS2008, WrightDM2015}. In particular, this association hosts two of the few known O3 stars in our Galaxy \citep{Walborn1973,WalbornHLM2002}, together with an incredibly luminous B supergiant, Cyg~OB2~\#12 \citep{MasseyThompson1991,NegueruelaMHC2008}. \par
  
  Together with this massive population, Cyg~OB2 is also rich in young low mass members. Censuses of low mass members based on $Chandra$/ACIS-I observations have found from 1000 to 1500 stars in the central region (down to $\sim 0.5\,$M$_{\odot}$, but complete at $\sim1\,$M$_{\odot}$; \citealp{AlbaceteColomboFMS2007} and \citealp{WrightDrake2009}). The age of these stars has been estimated to range between $3\,$Myrs and $5\,$Myrs by \citet{WrightDDV2010}, although a complex star formation history emerges from other studies: \citet{Hanson2003} dated some massive stars as younger than $2\,$Myrs; \citet{DrewGIS2008} identified a $5-7\,$Myrs old population of A stars southward the association center; stars with disks with intense accretion have been identified by \citet{VinkDSW2008}, and very young embedded protostars and active star forming regions are observed in the periphery of Cyg~OB2 \citep{WrightDDG2012, GuarcelloDWD2013}. In particular, the morphology of some of these structures, such as DR18 \citep{SchneiderBSJ2006}, with ongoing embedded star formation, and the observed orientation toward the center of Cyg~OB2 of several UV illuminated features \citep{Nicolainprep1}, clearly indicates a high level of feedback from the massive members of Cyg~OB2 \citep{GuarcelloDWD2013}. \par

Despite its proximity to the Sun, compared to other massive associations, Cygnus~OB2 is affected by high extinction mainly due to the dust associated with the Cygnus Rift in the foreground. Evidence that the Rift is responsible for an extinction of few magnitudes along this line of sight was first raised by \citet{DickelWendker1978}. \citet{SchneiderSBC2007} found an upper limit for the extinction due to the Rift equal to $A_V=5^m$. This was confirmed by \citet{SaleDUI2009}, who found an increase of extinction along this line of sight from $A_V=2^m$ to $A_V=5^m$ caused by dusts present in a region between 1 and $2\,kpc$ from the Sun. A similar estimate has been done by \citet{DrewGIS2008}, who found that the extinction affecting the stars associated with Cyg~OB2 ranges from $2.5^m<A_V<7^m$, and by \citet{WrightDDV2010}, who found a median $A_V=7.5^m$ in the central region and $A_V=5.5^m$ northward. A slightly lower extinction has been estimated by \citet{GuarcelloWDG2012} from $r-i$ vs. $i-z$ colors, finding a main range of extinction of $2.6^m<A_V<5.6^m$ with a median value $A_V=4.3^m$. The sample of candidate members of the association defined by \citet{Kashyapinprep} have larger extinctions, more similar to what has been found in previous studies: the 10\% and 90\% quantiles of their A$_V$ distribution are in fact $4.4^m$ and $8.5^m$, respectively, with a median value of $6.4^m$. \par

\begin{figure*}[]
        \centering
        \includegraphics[width=10cm]{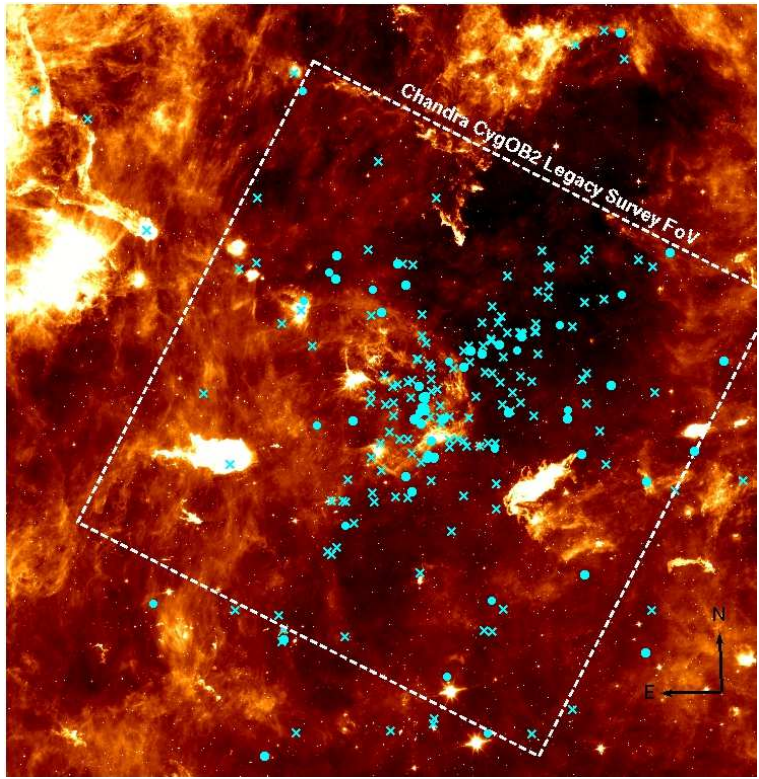}
        \caption{Spitzer/IRAC [8.0] image of the Cygnus~OB2 area. The field observed with ACIS-I is marked, as well as the positions of known O stars (filled circles) and B stars (crosses).}
        \label{field_img}
        \end{figure*}

  A region of 1 square degree centered on Cyg~OB2 has recently been surveyed with $Chandra$/ACIS-I for the $Chandra$ Cygnus~OB2 Legacy Project \citep{Drakeinprep}. The survey is composed of 36 ACIS-I fields overlapping each other in order to have a almost constant sensitivity in the central $40^{\prime} \times 40^{\prime}$ area. The resulting catalog of 7924 X-ray sources is described in \citet{WrightDGA2014}. Since optical and infrared data are crucial for most of the scientific aims of the survey, the X-ray catalog has been combined with a large set of photometric data: optical data in $riz$ bands from specific observations with OSIRIS@GTC \citep{GuarcelloWDG2012} down to $r=25^m$ ; data in $ugriz$ bands from the SDSS/DR9 public catalog \citep{AiharaAAA2011} down to $r=16^m$; data in $r^{\prime}i^{\prime}H_{\alpha}$ bands from the second data release of the IPHAS public catalog \citep{DrewGIA2005,BarentsenFDG2014} down to $r^{\prime}=21.5^m$; NIR data in $JHK$ from 2MASS/PSC \citep{CutriSDB2003} and UKIDSS/GPS \citep{LucasHLS2008, KingNBG2013}, down to $J=18.5^m$ and $J=21^m$, respectively; Spitzer and MIPS data from the {\it Spitzer Legacy Survey of the Cygnus X region} \citep{BeererKHG2010}. With the exception of the OSIRIS catalog, covering the central $40^{\prime}\times40^{\prime}$ area, all these catalogs cover the entire survey area. This large set of photometric data has been combined in an unique optical-infrared-X-ray catalog containing 328540 sources (\citealp{GuarcelloDWD2013}, and 2014). Fig. \ref{field_img} shows a $8.0\mu$m image of the region of Cyg~OB2, with indicated the area observed with $Chandra$/ACIS-I, together with the positions of known O and B stars. \par
  
\begin{figure*}[]
        \centering
        \includegraphics[width=8.0cm]{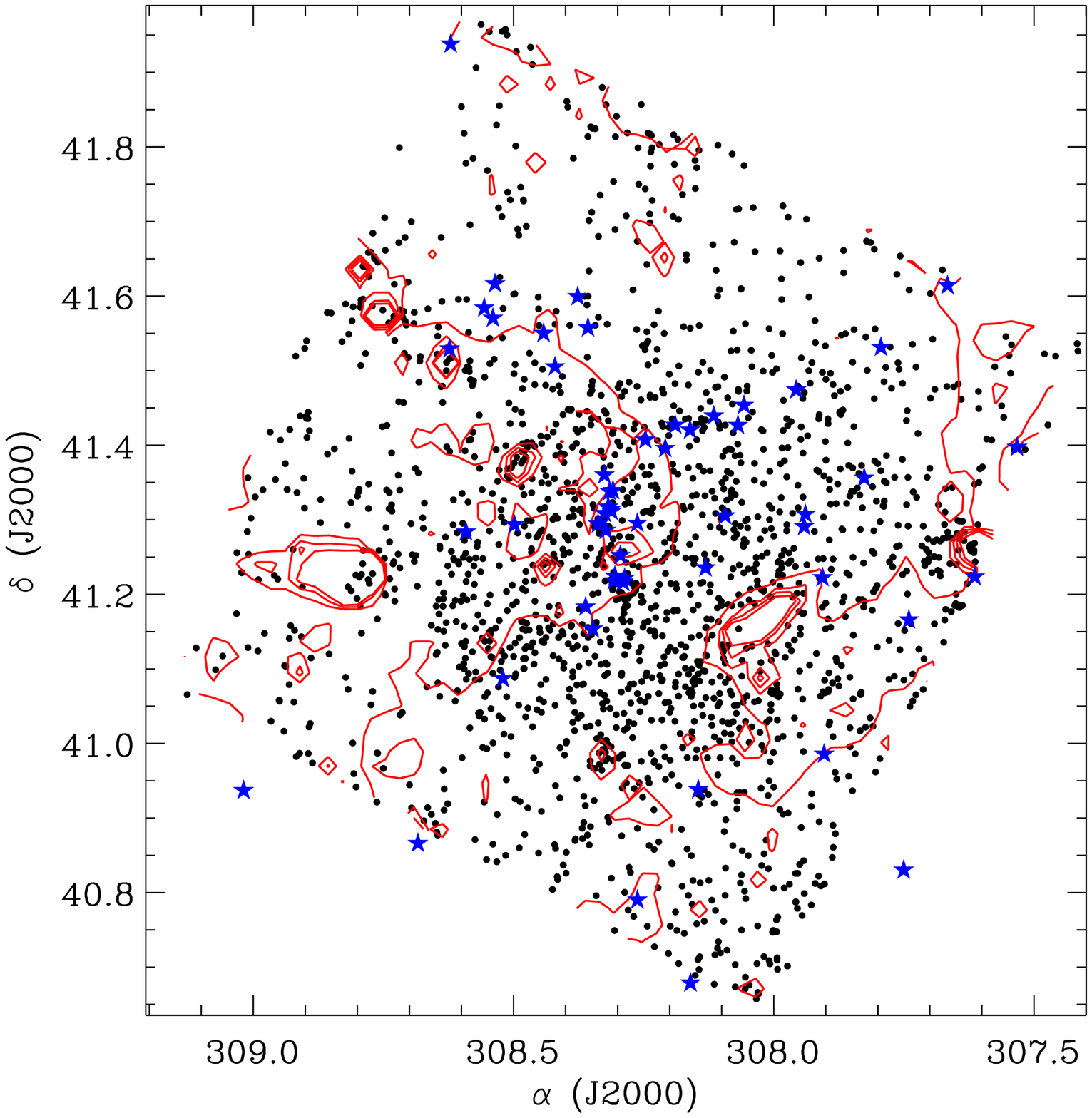}
        \includegraphics[width=8.0cm]{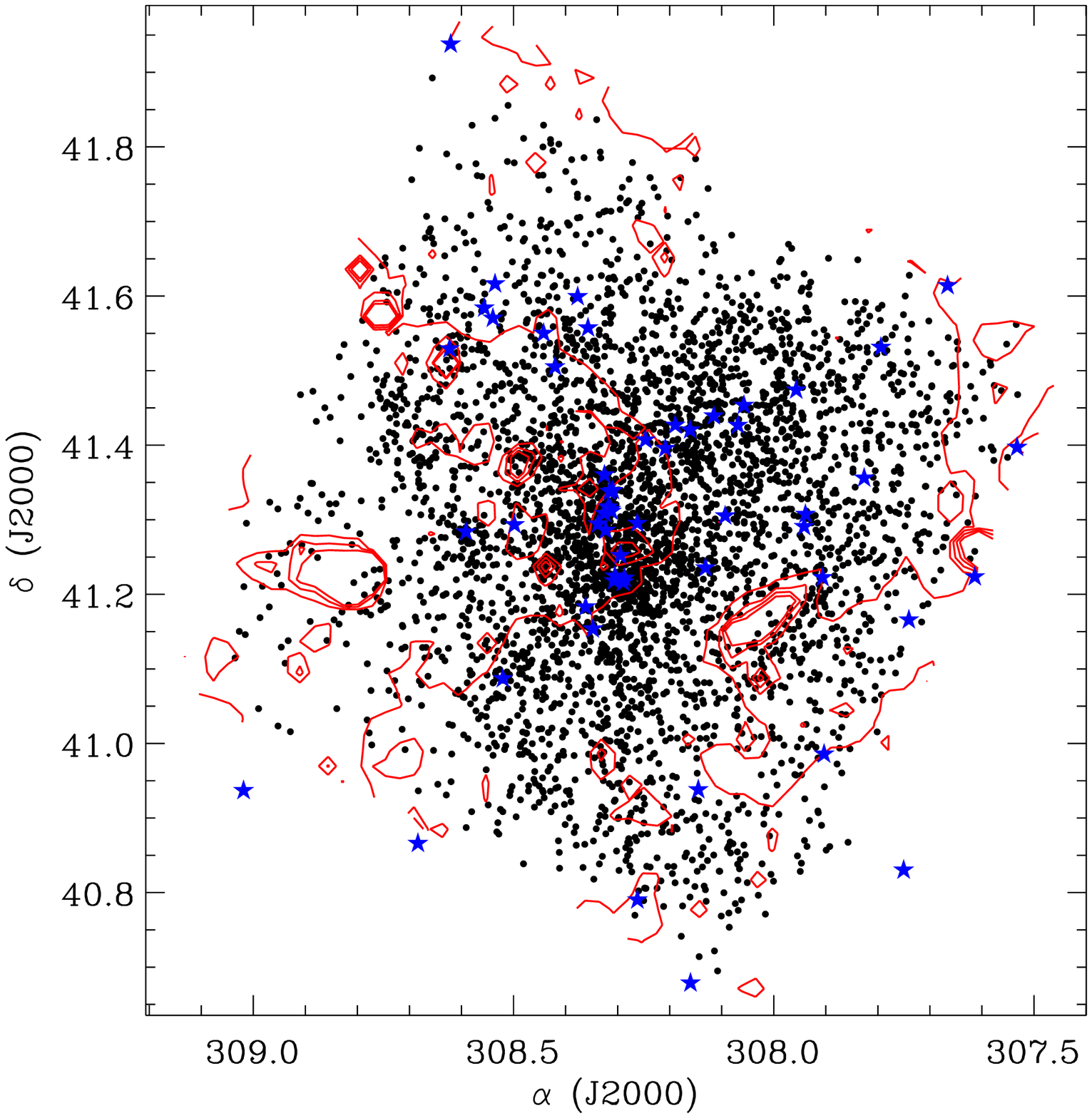}
        \caption{Spatial distribution of stars with disks (left) and without disks (right), showing the position of the O stars (blue star symbols) and the $8\,\mu$m continuum emission contours in red (color figure in the electronic paper).}
        \label{spadis_img}
        \end{figure*}

In this paper we study how the disk fraction (i.e.: the fraction of stars associated with Cyg~OB2 bearing a circumstellar disk) changes across the $Chandra$ Cygnus~OB2 Legacy Survey area as a function of the local UV (FUV and EUV) field and the local surface stellar density. For this aim, we use the list of 1843 members with disks selected by \citet{GuarcelloDWD2013}, which has been purged from contamination by foreground field stars, background giants, and extragalactic sources; and the list of 5022 candidate members obtained by \citet{Kashyapinprep} from the catalog of X-ray sources described in \citet{WrightDGA2014}. Among these X-ray sources, 441 are disk-bearing stars and 102 are known OB stars. The remaining 4479 (with 154 multiple matches between the X-ray and optical-infrared catalogs, \citealp{GuarcelloDWN2015}) are good candidate class~III objects of the association. Fig. \ref{spadis_img} shows the spatial distribution of the selected members with disks (left panel) and without disks (right panel), overplotted with the position of the O stars. The contours mark the emission levels at $8.0\mu m$ from Spitzer observations, revealing the dense nebular structures. Some of these are sites of ongoing star formation, such as DR18. In the spatial distribution of the disk-bearing sources, it is possible to distinguish the central cluster and the surrounding annular stellar overdensity discussed in \citet{GuarcelloDWD2013}. Also the spatial distribution of the members without disks shows a clear overdensity corresponding to the central cluster, as well as a rich population in the outer regions. \par
  
  For the purposes of this paper, it is necessary to estimate the UV flux emitted by the O stars associated with Cyg~OB2. We will omit from the calculation the flux emitted by the early-B stars, whose census is still incomplete. Table \ref{ostars_tb} shows the positions and spectral types of the known O stars, together with their expected FUV and EUV luminosity. Their FUV luminosity have been obtained using the interpolated relations from \citet{ParravanoHM2003}. Stellar masses and spectral types, necessary to calculate the FUV luminosity, are taken from \citet{WrightDM2015}. The EUV luminosity in Table \ref{ostars_tb} are in units of number of ionizing photons with $\lambda < 912\,$\AA{} per second (the $Q_0$ value in \citealp{MartinSH2005}). We included in the list the three WR stars of the association and the O stars which are outside the field of our survey but within a few arcminutes. \par

\section{Spatial variation of the disk fraction in Cygnus~OB2}
\label{df_sect}

\begin{figure*}[]
        \centering
        \includegraphics[width=15cm]{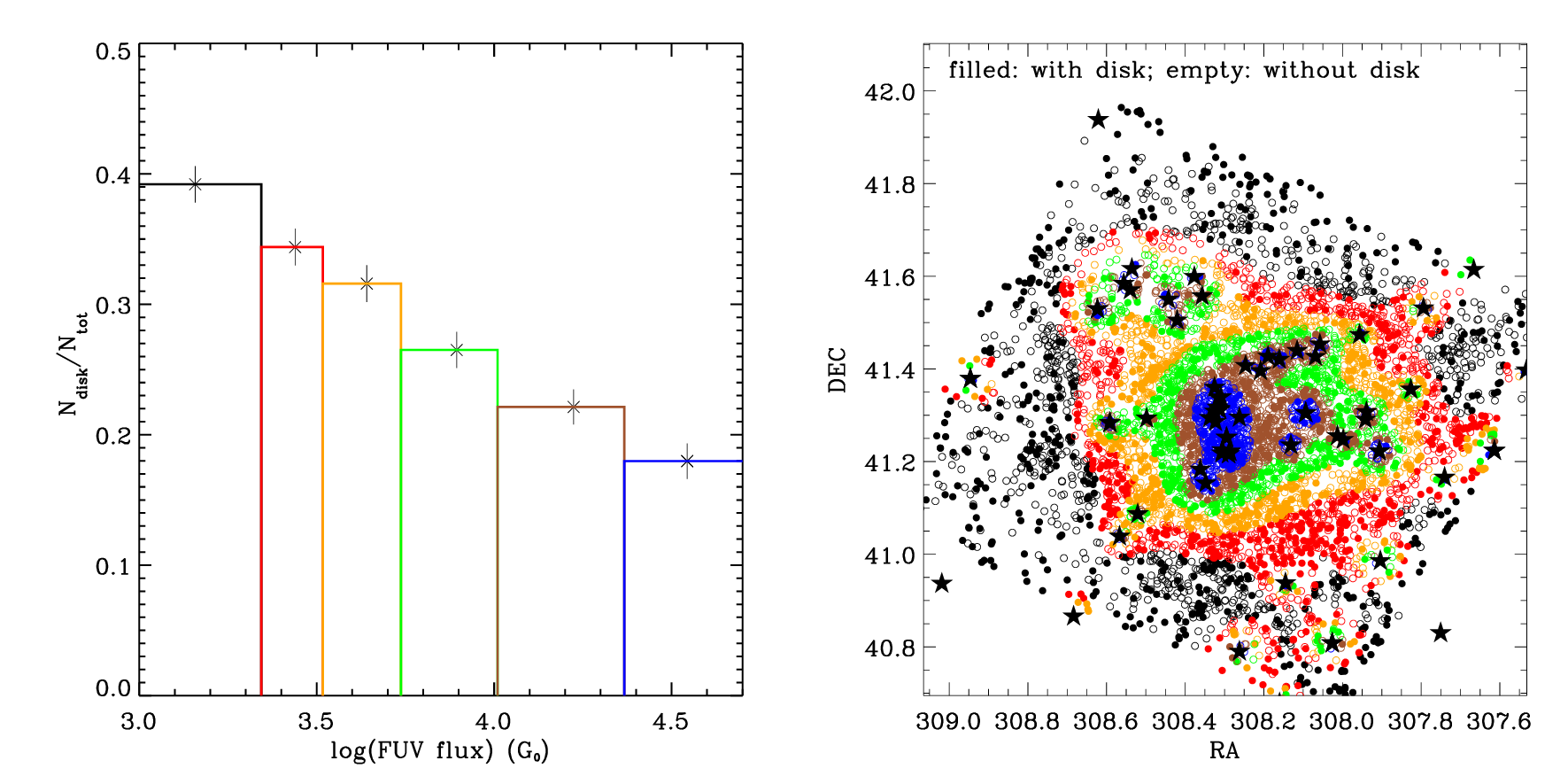}
        \includegraphics[width=15cm]{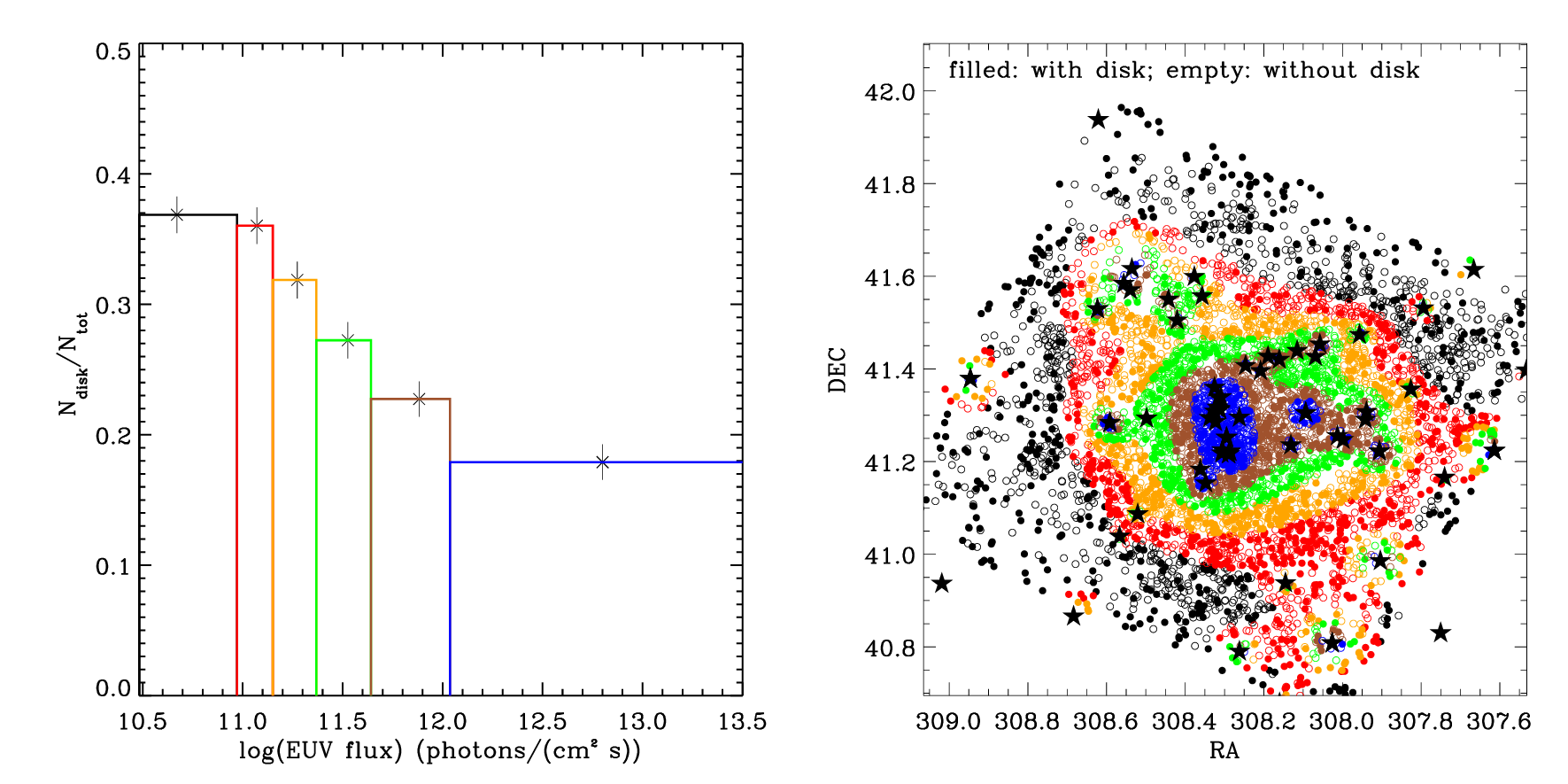}
        \caption{Variation of the disk fraction as a function of the local EUV (bottom left panel) and FUV fields (top left panel) experienced by the low-mass members. The spatial distribution of the stars falling in each bin is shown in the right panels, with the different colors marking those stars falling in the corresponding bin in the left panels. The star symbols mark the positions of the O stars.}
        \label{dfuv_img}
        \end{figure*}

To study the effect that O stars have on disk evolution in Cyg~OB2, we need to calculate how the disk fraction varies across the association as a function of the local EUV and FUV radiation fields. For this aim, using the technique adopted in \citet{GuarcelloPMD2007}, we calculate the FUV and EUV fluxes emitted by each O star and incident at the position of each star associated with Cyg~OB2, both with and without a disk, using the projected distances (i.e. the 2D projection of the real distances). In this way it is possible to calculate the disk fraction in given ranges of incident UV flux. The results are shown in Fig. \ref{dfuv_img}. The upper and lower left panels show the disk fraction variation vs. the incident FUV and EUV fluxes, respectively. The FUV fluxes are measured in terms of the Habing flux $G_0$, equal to $1.6\times10^{-3}\,$erg/cm$^{2}$/s; 1.7$\,$G$_0$ corresponds to the average UV flux in the $912-2000\,$\AA{} spectral range in the Solar neighborhood \citep{Habing1968}. The EUV fluxes are described in units of photons/s/cm$^{2}$. The size of the bins is fixed in order to include the same number of disk-less members in each of them, so that the observed variation is given by the decrease of the number of stars with disks. The upper limit of incident fluxes used in Fig. \ref{dfuv_img} (i.e. log$\left( F_{FUV} \right)=4.7$ and log$\left(F_{EUV}\right)=13.5$) are chosen only to improve the visualization of the figure and they are much smaller than the actual upper limits of the flux incident on the stars. The right panels show the spatial distribution of the candidate members (both with and without disks) and the O stars. The different colors are used to mark the position of the stars falling in each of the bins in the right panels. \par
  
    Fig. \ref{dfuv_img} clearly shows a smooth decline of the disk fraction from $\sim40\%$ in the regions with low UV fluxes to $\sim18\%$ in those close to the O stars.  Hereafter, the bins will be labeled with recurring numbers starting from those at high fluxes. In this way, the first bin is the one corresponding to the highest UV fluxes, the sixth bin the one corresponding to the lowest fluxes. \par

In this calculation, we are not accounting for any evolution of the UV radiation field in Cyg~OB2 due to the evolution of the O stars. This approximation should not have a strong impact on our calculation given the age of Cyg~OB2 members and the expected evolutionary timescale of these massive stars. Star formation in Cyg~OB2 is expected to have started about 5$\,$Myrs ago \citep{WrightDDV2010}, with most of the members formed about 2-3$\,$Myrs ago \citep{MasseyDW2001}. Since:  1) The lifetime of O stars is expected to range between 4$\,$Myrs and 7$\,$Myrs for the early \citep{LangerHLN1994} and late \citep{MeynetMSS1994,SchaererDeKoter1997} O stars, respectively; 2) star formation is ongoing only in the periphery of the association \citep{GuarcelloDWD2013}; the use of present-day UV flux is a good approximation of the overall UV field experienced by the disks during the evolution of the association. It must also be noted that the study of \citet{WrightPGD2014} showed that in the past Cyg~OB2 never had a highly clustered stellar configuration. {\bf We are also ignoring for the moment the absorption of the UV radiation by intracluster dust particles that may be still present in the association. Given that even reasonably small concentrations of dust particles may result in significant attenuation of the UV radiation, this may play an important role in shaping the disk fraction vs. incident UV flux relation we observe, especially in the case of disks at large distance from the center of the association. The effects of this absorption are described in more detail in Sect. \ref{safe_sec}.}\par
  
  The spatial variation of disk fraction shown in Fig. \ref{dfuv_img} is compatible with a scenario where disks closer to O stars are dissipated faster by externally induced photoevaporation than those in the outer part of the association. However, there are other effects that can produce similar results: a sequence of star formation from the center of the association outward such that the central stars are significantly older, a 2D projection effect and a non uniform sensitivity of the data sets employed to identify Cyg OB2 members with and without disks. \par

\subsection{2D projection effect}
\label{2d_sec}

\begin{figure}[]
        \centering
        \includegraphics[width=11cm]{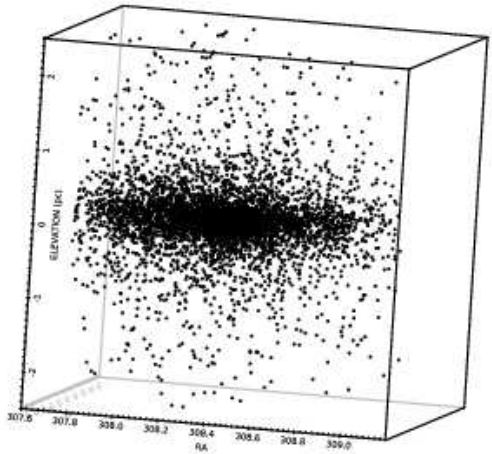}
        \caption{Simulated 3D stellar spatial distribution of the association.}
        \label{3d_img}
        \end{figure}

\begin{figure}[]
        \centering
        \includegraphics[width=8.5cm]{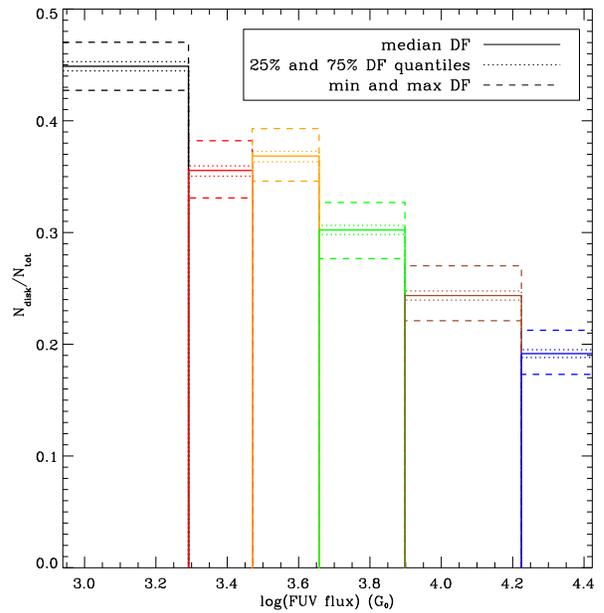}
        \caption{Variation of the disk fraction as a function of the local FUV radiation field obtained in the 5000 simulated 3D configurations of the association. The solid histogram is obtained from the median disk fractions observed in the FUV bins; the dotted histograms from the 25\% and 75\% quantiles, while the dashed lines show the minimum and maximum disk fractions observed in each FUV bin.}
        \label{3d_histo}
        \end{figure}

  {\bf In Sect. \ref{df_sect} we describe how we calculate the FUV and EUV fluxes at the position of the candidate members using the projected distances between each member to the O stars. The projected distances are lower limits to the real distances between stars, and they lead to overestimating the UV fluxes experienced by the candidate members. Since we are studying the effects on the disk fraction, i.e. the ratio of the number of stars with disks over the total number of members, it is not trivial to deduce how the 2D projection of the real stellar distribution affects the correlation observed between disk fraction and incident UV flux. One way to infer how the trend in Fig. \ref{dfuv_img} would appear using the real 3D distances between stars is by simulating a large number of realistic morphologies of the association and calculating the disk fraction vs. FUV flux for each simulation. We want to verify whether possible 3D configurations of the association exist where the decrease of the disk fraction as a function of the incident UV flux would not be observed. \par

	In order to simulate the 3D morphology of the association, we need to assign to each member an {\it elevation} ({\it z}, in parsec) from the 2D projection plane. The elevation can be either positive or negative. If Cyg~OB2 were a cluster and not an association, it would have been appropriate to adopt as an upper limit of $z$ a multiple (such as 5 times) of the cluster radius. Cyg~OB2 is, however, an association containing several subclusters and structures (e.g. \citealp{GuarcelloDWD2013}), and Cyg OB2 has been shown to have a substructured and fractal structure and is not completely mixed \citep{WrightPGD2014}, so that stars that are close together on the plane of the sky will also be close together along the line of sight. More realistic simulated 3D configurations can be obtained by considering its clumpy structure. \par
 
We first calculate the {\it Minimum Spanning Tree} (MST) of the members \citep{BarrowBS1985} using the $R$ statistics package \emph{nnclust}\footnote{http://cran.r-project.org/web/packages/nnclust/nnclust.pdf}. The MST is defined as the unique set of branches connecting all the points in a given data set with the minimum total length and not producing closed loops. This technique is typically used to select and extract clustered groups of points, and it has been used in several different studies related to stellar clusters (e.g. \citealt{GutermuthMMA2008}). In this case we do not attempt any selection of subclusters, but instead we simulate an isotropic distribution for the orientation of each branch of the MST with respect to the projection plane adopting a uniform probability distribution. We set an upper limit to the simulated elevation of each star equal to $z_{max}=\pm 20\,$pc, in order to avoid elevations ranging from 0 to $\pm \infty$. \par
  
We run 5000 simulations. Fig. \ref{3d_img} shows one of the 3D distributions that we obtain. A significant stellar population at large elevations apparent in Fig. \ref{3d_img} is found to be a common feature of all the simulated 3D configurations. However, the fall-off of the distribution towards the upper and lower box edges shows that out result is not unduly sensitive to the value of z$_{max}$ chosen. \par
  
  For each simulation, we calculate the disk fraction variation as a function of the incident FUV flux, using the simulated 3D distances between the low mass members and the O stars. The decline of the disk fraction toward the O stars is always observed. In Fig. \ref{3d_histo} we show the median and 25\% and 75\% quartiles of the disk fractions as a function of UV flux for all the realizations. The difference between the 25\% and 75\% quantiles of the disk fraction shown in Fig. \ref{3d_histo} is similar to its error bars as shown in Fig. \ref{dfuv_img}, meaning that the 2D approximation has a very small impact on our study. In conclusion, there is no evidence that the variation of the disk fraction as a function of the incident UV flux observed in Cyg~OB2 is a consequence of the 2D projection.} \par

\subsection{The sequence of star formation}
\label{trig_sec}

\citet{Mamajek2009} describes the decline of disk fraction with cluster age as an exponential decay with an e-folding timescale of $2.5\,$Myrs. As these authors state, however, the observed decay strongly depends on the diagnostics adopted to select members with and without disks and on the type of disks included in the selection (i.e., active, passive, transitional, etc...). The disk selection in Cyg~OB2 has been performed with $JHK$, Spitzer/IRAC, and Spitzer/MIPS 24$\,\mu$m data, and it involves both stars with thick disks, candidate stars with transition, pre-transition, and low-mass (anemic) disks \citep{GuarcelloDWD2013}, classified according to their spectral energy distributions and infrared colors. \par

  A decrease of disk fraction from $\sim40\%$ to $\sim18\%$ as we observe in Fig. \ref{dfuv_img} can be explained if the population close to the O stars is about $5\,$Myrs old, while that in the outer part of the association is younger than $2\,$Myrs. Such a chronology can be a consequence of a triggered star formation in the periphery of the association by the OB stars in the center, which is a plausible hypothesis in Cyg~OB2. In order to examine this possibility, it is necessary to estimate the age of the stars in the different bins in Fig. \ref{dfuv_img}. \par

  A rigorous estimate of stellar age would require an accurate spectral classification of all the members, for which spectroscopic data are still not available. A less accurate method, based on multi-band photometry that is available for most of the stars in Cyg~OB2, consists in interpolating the position of each member in dereddened optical color-magnitude diagrams with the pre-main sequence isochrones from \citet{SiessDF2000} and the PARSEC stellar evolutionary model \citep{BressanMGS2012}\footnote{Downloaded from the CMD 2.7 web interface}. Since the presence of a circumstellar disk can alter the optical colors of the central star by adding a blue-excess due to the accretion process \citep{CalvetGullbring1998} or scattering of photospheric optical radiation into the line of sight \citep{GuarcelloDMP2010}, this method for evaluating stellar age can not be applied to the disk-bearing members but only to candidate class~III objects.\par

The optical $r$ vs. $r-i$ and $r-i$ vs. $i-z$ diagrams of all the X-ray sources with optical counterparts are shown in \citet{GuarcelloWDG2012}. Most of these sources are classified as members of Cyg~OB2, so these diagrams also show the loci of the disk-less members. Since the extinction in Cyg~OB2 changes drastically with position, in order to obtain a reliable estimate of the age of the disk-less members it is necessary to deredden their optical colors using an estimate of the individual extinction of each star, rather than some approximated average value. \citet{GuarcelloWDG2012} used the displacement along the reddening vector from the $A_V=0^m$ $2.5\,$Myrs isochrone in the $r-i$ vs. $i-z$ diagram to calculate the individual extinctions of the optical+X-ray sources in the central region of the association. In the $Chandra$ Cygnus~OB2 Legacy Survey we adopt the same approach but using the main sequence locus defined by \citep{CoveyISF07} rather than those from \citet{SiessDF2000} and the source classification provided by \citet{Kashyapinprep}, calculating ages and masses only for the association members. \par

\begin{figure*}[]
        \centering
        \includegraphics[width=14cm]{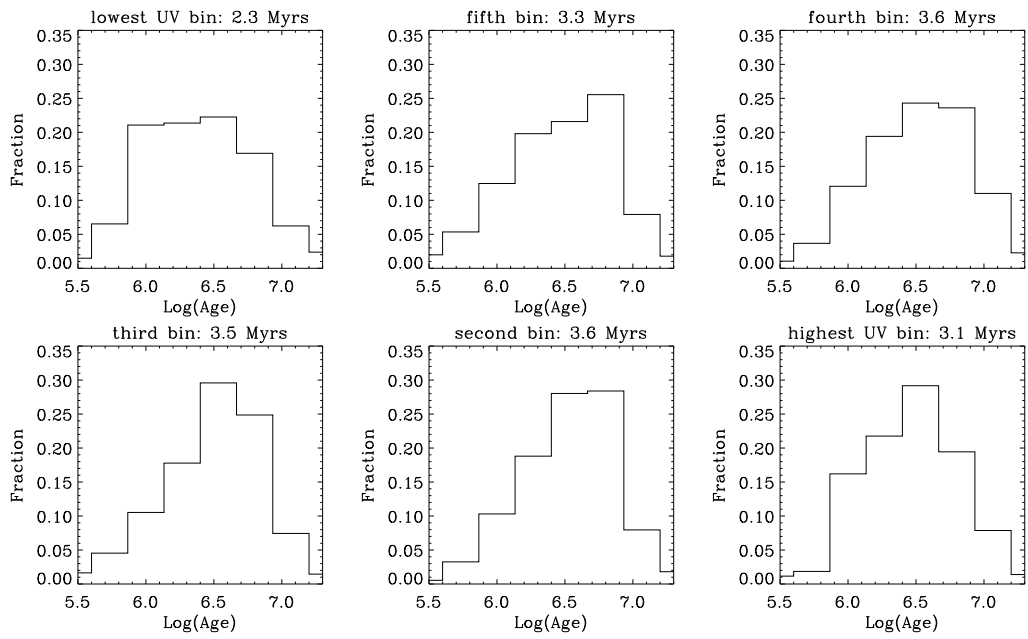}
        \caption{Each histogram represents the age distribution of the stars falling in the bins in Fig. \ref{dfuv_img}. The plot titles show the median age of each distribution.}
        \label{age_img}
        \end{figure*}

  Fig. \ref{age_img} shows the age distributions calculated with \citet{SiessDF2000} isochrones of the disk-less stars falling in each of the six bins of Fig. \ref{dfuv_img}. The main result is that the distribution of stellar ages for the stars falling in the first and last bin is similar, with a difference of median age of only 0.8$\,$Myrs. This age difference, as well as the difference observed among the stars falling the other bins, can not account for the observed difference of disk fraction. 
A similar result is found using the PARSEC models (in this case the difference of median age is $0.6\,$Myrs). In both cases, only the stellar population of the sixth bin shows a wider distribution, with a population of younger stars falling in the active star forming regions identified by \citet{GuarcelloDWD2013}. This finding is in agreement with the lack of significant spatial variation of stellar age as found by \citet{WrightDM2015} for the OB population. \par
  
As a further test, we fit the $r$ and $r-i$ distributions of the populations of the six bins with a set of isochrones using the maximum likelihood $\tau^2$-minimization method \citep{NaylorJeffries2006,Naylor2009}. This method employs fitting of two-dimensional photometric data in a color-magnitude space, using a wide set of isochrones and also taking into account the effects of the binary population as well as observed photometric errors. Since the method does not calculate the extinction separately, we use the deredden colors as described above. We fit the $r_0$ vs. $r_0-i_0$ distribution of the stellar population of the six bins using all the available isochrones \citep{DAntonaMAzzitelli1997,BaraffeCAH1998,PallaStahler1999,SiessDF2000,Dotter2008,TognelliDP2012}, adopting a binary fraction of 0.5 and solar metallicity. Since we correct for the individual reddening, we do not apply any further systematic color error. Only when using the isochrones from \citet{Dotter2008} we observe a likely inside-out age gradient with a difference of 3$\,$Myrs between the inner and outer stellar populations. With the remaining six sets of isochrones we use, no such a trend is observed, with the age difference between the stellar populations of adjacent bins being typically smaller than 1$\,$Myr. Assuming that the disk-bearing and disk-less population are almost coeval, we can conclude that the observed decline of disk fraction toward the O stars in Cyg~OB2 is not a consequence of an inside-out sequence of star formation.
  
\subsection{Completeness}
\label{comp_sec}

Any not uniform sensitivity of our survey across the field might impact the spatial variation of disk fraction that we observe. For instance, the sensitivity across the ACIS-I detector is known to decrease with the displacement from the center. Since pre-main sequence stars are in general in the saturated activity regime \citep{PreibischFeigelson2005}, which means that the X-ray luminosity in these stars is independent from stellar rotation but it scales with bolometric luminosity, which itself is dependent on stellar mass and age, single ACIS-I observations of young clusters will be deeper in stellar mass at small off-axis angles than in the outer part of the field of view. {\bf Since there are indications that less massive stars dissipate their disks more slowly than more massive stars \citep{CarpenterMHM2006ApJ,CarpenterMHM2009ApJ,LuhmanMamajek2012ApJ}}, and also accounting for the fact that disk-bearing stars are in general less X-ray bright than disk-less stars \citep{PreibischKFF2005,FlaccomioMS2003}, this can affect the observed spatial variation of disk fraction. However, this effect would mainly tend to increase the observed disk fraction toward the center of a single field. Even if the $Chandra$ Cygnus OB2 Legacy Survey was designed in order to reduce the impact of not uniform sensitivity \citep{Drakeinprep,WrightDG2015}, it is important to verify whether this issue may impact our results.\par

  Another important effect can be the loss of sensitivity in the optical and infrared images close to very bright stars, such as most of the O stars in Cyg~OB2. The bright wings of the PSF of such stars can result in brighter background and larger difficulty in detecting the emission from nearby faint stars. It is not easy to predict the impact that this problem has in our study, since observations at different wavelength are affected in different ways. \par
  
About 95\% of the candidate cluster members have $JHK$ counterparts, in general from UKIDSS. The simplest way to verify whether a not uniform sensitivity of the data has affected the result shown in Fig. \ref{dfuv_img} is to derive the variation of the disk fraction as a function of the FUV flux with different cuts in $J$ magnitude. Considering only members with $J\leq 17^m$, $J\leq 16^m$, or $J\leq 15^m$ the disk fraction declines from 35\% to 19\% in the former two cases and from 44\% to 24\% in the latter. We run further tests to verify whether other completeness issues may affect our results. We observe a smooth decline of the disk fraction toward the center even considering only candidate class~III sources brighter in X-rays than the 25\%, 33\%, and 50\% quantile of the X-ray fluxes distribution, with a total decline of the disk fraction from the 46\% to 23\%, 49\% to 25\%, and 62\% to 33\% respectively. The masses of the candidate cluster members detected in X-rays are estimated from the dereddened optical and infrared color-magnitude diagrams \citep{Kashyapinprep}. Even if the presence of the disk can affect the optical colors observed from the class~II sources, and thus the estimate of the individual extinction and mass, we adopt the same procedure to estimate the masses of the disk-bearing stars not detected in X-rays, and calculate the disk fraction vs. the incident FUV flux adopting two cuts in stellar mass: 0.4$\,$M$_{\odot}$ and 0.7$\,$M$_{\odot}$. The resulting spatial variation of the disk fraction is still characterized by a significant decrease toward the center of the association, even if with a smaller total disk fraction given the small number of stars for which masses can be calculated: from 21\% to 15\% and from 25\% to 17\% respectively. \par

	{\bf As a last test, disk lifetimes may depend on the mass of the central star \citep[e.g. ][]{CarpenterMHM2006ApJ,CarpenterMHM2009ApJ,LuhmanMamajek2012ApJ}}. If the mass content of the central cluster is different than that of the population in the outskirt of our field, this would affect the spatial variation of the disk fraction. However, the disk fraction calculated only for members with M$\leq 1\,$M$_{\odot}$ declines from 27\% to 13\%, while for members with M$\leq 0.6\,$M$_{\odot}$ from 28\% to 15\%. Analogously, the disk fraction for members with $J\geq15.4/m$ or $J\geq16.1^m$ (i.e. the expected $J$ magnitudes of a 1$\,$M$_{\odot}$ and 0.6$\,$M$_{\odot}$ stars, respectively, at 1400$\,$pc of distance and extinguished by A$_V$=4.5$^m$) declines from 39\% to 18.5\% and from 44\% to 21\%, respectively. 

\section{Photoevaporation vs. disruption due to close encounters}
\label{map_sec}

In this section we compare the effects on disk evolution due to UV radiation and close encounters adopting two different approaches: the analysis of disk fraction maps with a higher spatial resolution than employed in Fig. \ref{dfuv_img}; and estimating the local stellar density around the position of each member simulated in the 15000 simulations of the 3D configuration of the association as discussed in Sect. \ref{2d_sec}.

	\subsection{Disk fraction maps}
	\label{dfmaps_sec}

\begin{figure*}[]
        \centering
        \includegraphics[width=10.0cm]{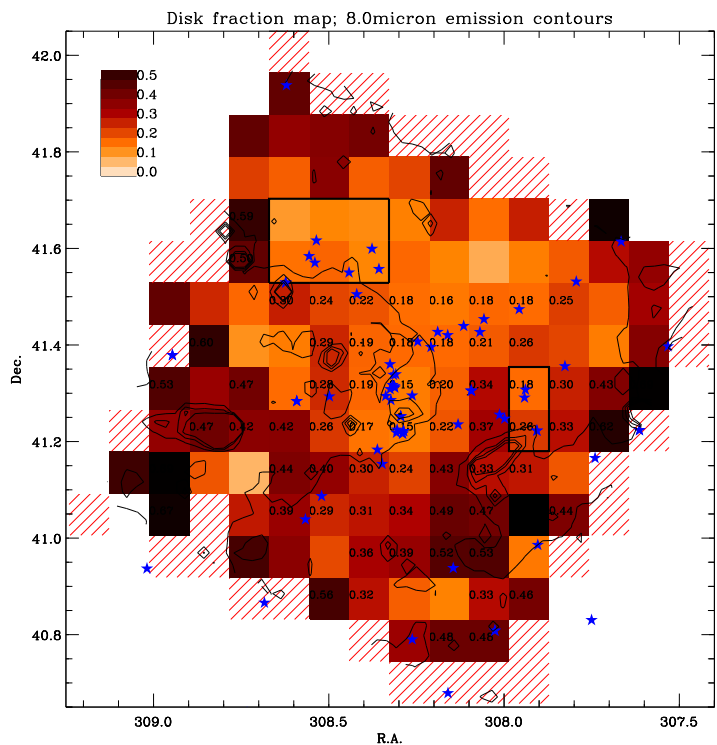}
        \caption{Maps showing the spatial variation of the disk fraction coded with red tones. The values are indicated only in those bins where the relative error of the disk fraction is smaller than 0.25. The overplotted contours mark the $8.0\, \mu m$ emission.}
        \label{dfmapirimg}
        \end{figure*}

Fig. \ref{dfmapirimg} shows a map of the disk fraction calculated in a grid of $15\times 15$ bins and encoded with red tones. The value of the disk fraction in those bins where $\sigma_{DF}/DF<0.25$ is overplotted ($DF$ indicates the disk fraction).  \par

  In Fig. \ref{dfmapirimg} the contours mark the emission intensity at $8.0\mu m$ as inferred from Spitzer data. This map allows us to identify those bins where local peaks of the disk fraction correspond with nebular structures that have been identified as active star forming regions in \citet{GuarcelloDWD2013}, such as the DR18 cloud at $\alpha \sim 308.8,\, \delta \sim 41.2$, the {\it Globule 1} at $\alpha \sim 308.0,\, \delta \sim 41.2$ \citep{SchneiderBSJ2006} and the {\it bright rimmed clouds} in the north-east from the central cluster\footnote{Approximately at $\alpha \sim 308.6,\, \delta \sim 41.5$, $\alpha \sim 308.7,\, \delta \sim 41.55$, and $\alpha \sim 308.8,\, \delta \sim 41.6$} \citep{GuarcelloDWD2013}. On average, the disk fraction in these structures ranges from $\sim10\%$ to $\sim20\%$ larger than that in some of the surrounding bins. This is compatible with the embedded stellar population being $1.5-2\,$Myrs younger than the surrounding stars \citep{Mamajek2009}. \par

\begin{figure*}[]
        \centering
        \includegraphics[width=8.0cm]{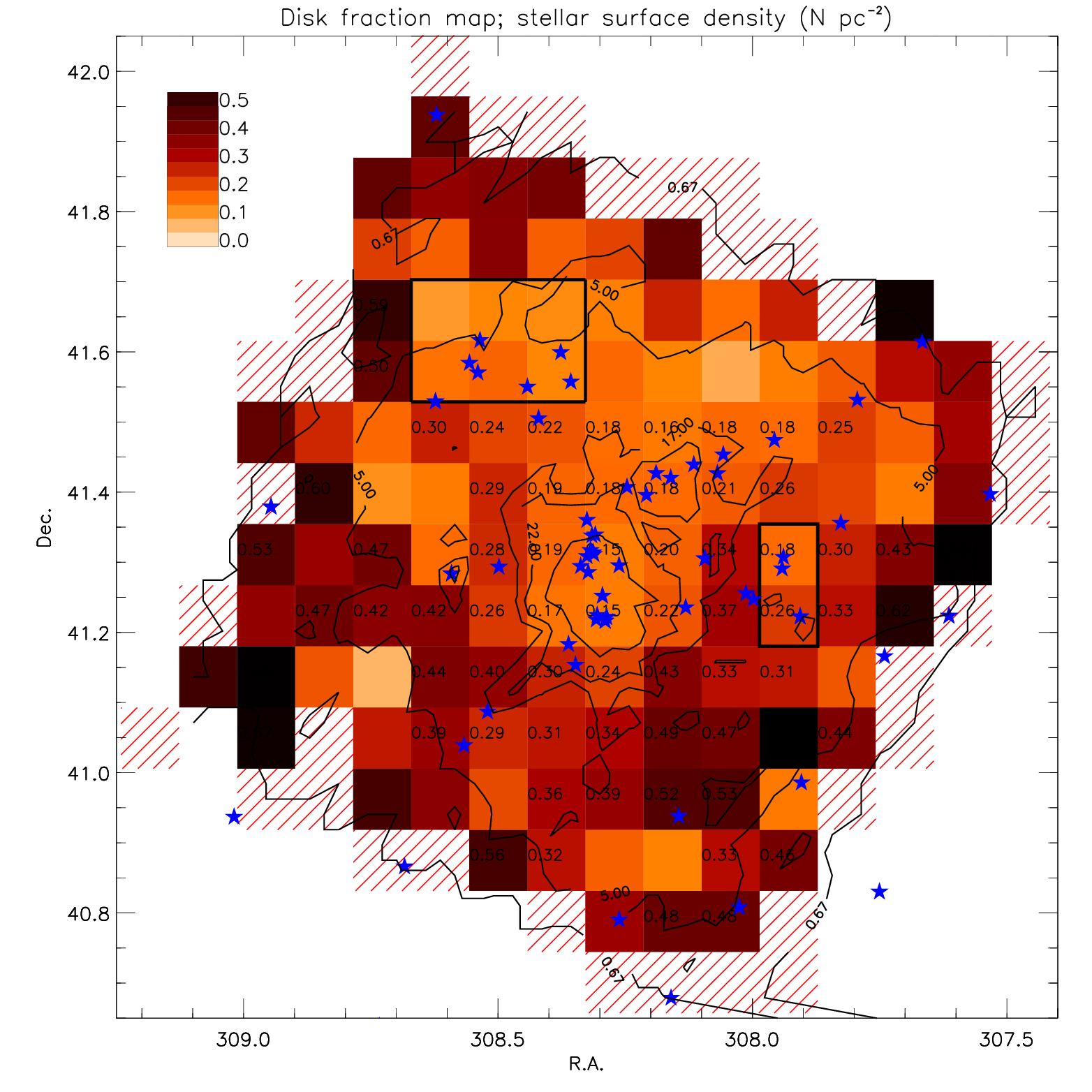}
        \includegraphics[width=8.4cm]{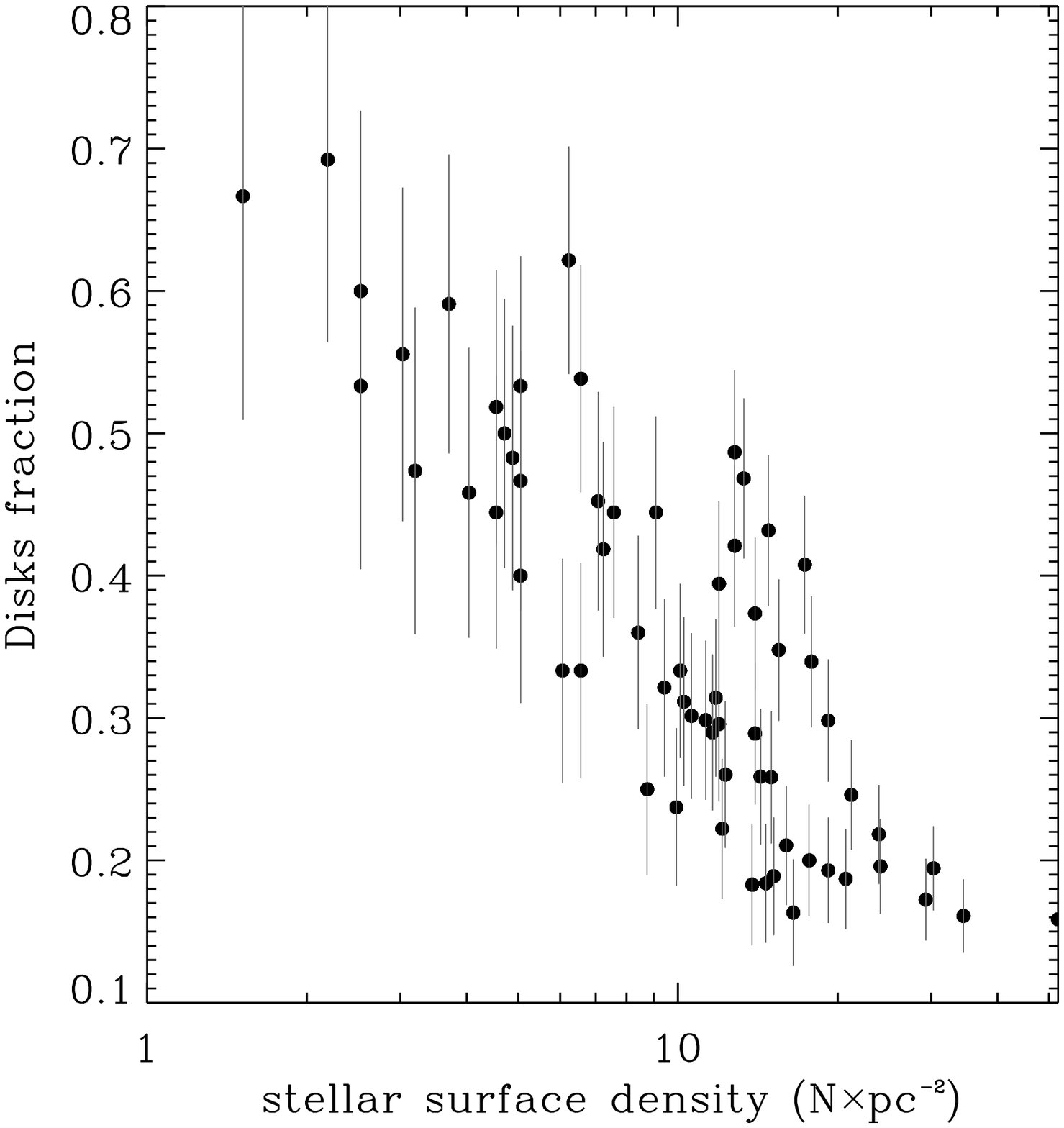}
        \caption{Left panel: Maps showing the spatial variation of the disk fraction as in Fig. \ref{dfmapirimg}. The overplotted contours mark the local stellar surface density. The contour levels are: 0.7, 5, 17, 22, and 33 N/pc$^{2}$. Right panel: Disk fraction vs. stellar surface density flux in the bins of the same grid, showing only the values measured in bins with the relative error of the disk fraction smaller than 0.25.}
        \label{dfmapdensimg}
        \end{figure*}

In the left panel of Fig. \ref{dfmapdensimg} the spatial variation of disk fraction is compared with the stellar surface density. Contours mark the area where the local stellar surface density is equal to typical values of in clusters within 2$\,$kpc from the Sun. The adopted limits are the 25\% and 95\% quantiles of the distribution of surface densities of the clusters within $2\,$kpc from the Sun, as compiled by \citet{LadaLada2003} and \citet{PorrasCAF2003}: $18\,$N$\,$pc$^{-2}$ and $22\,$N$\,$pc$^{-2}$, respectively. The limit corresponding to the densest region in Fig. \ref{dfmapdensimg} is $33\,$N$\,$pc$^{-2}$, which is the slope of the $N$ vs. $R$ relation for these clusters (with $R$ being the cluster radius) as found by \citet{AdamsPFM2006}. These low  densities are in agreement with the finding of \citet{WrightPGD2014} who have demonstrated that Cyg~OB2 has always been an association, characterized by modest stellar density such as that we observe. The right panel shows clear decrease of the disk fraction as a function of the stellar surface density measured in those bins with low relative error in the disk fraction. \par

\begin{figure*}[]
        \centering
	\includegraphics[width=8.0cm]{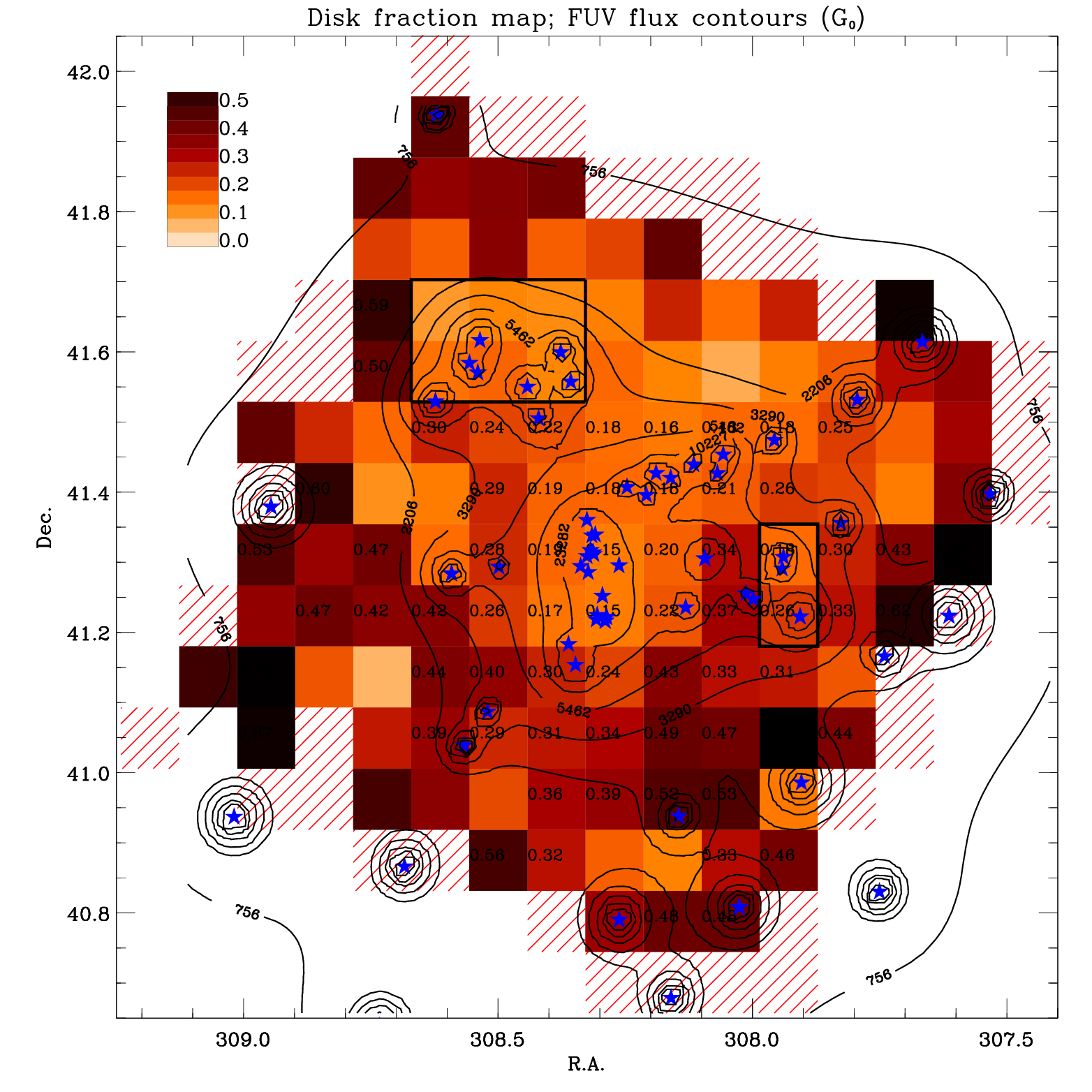}
        \includegraphics[width=8.4cm]{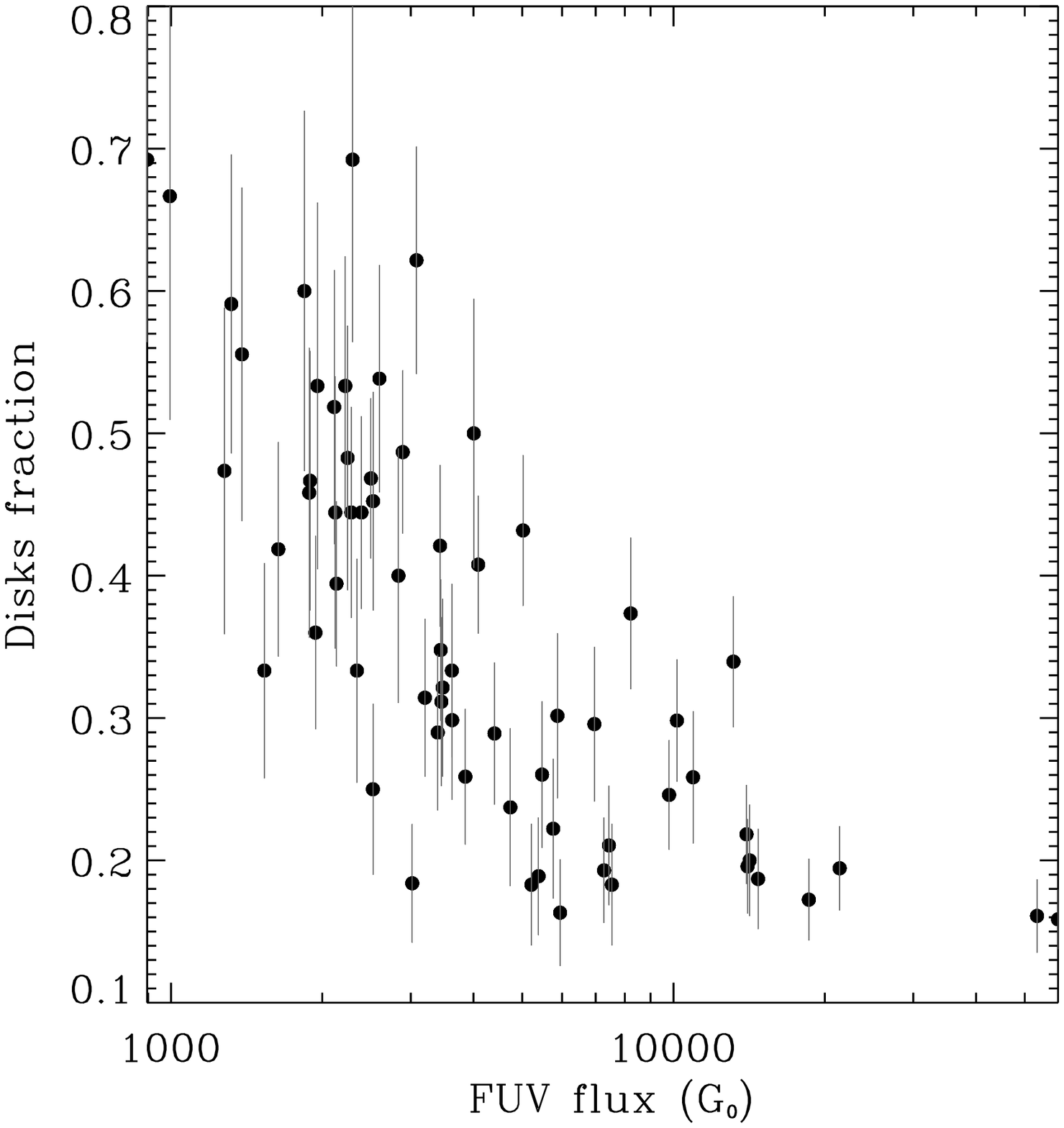}
        \caption{Left panel: Maps showing the spatial variation of the disk fraction as in Fig. \ref{dfmapirimg}. The overplotted contours mark the local FUV flux. The contour levels are: 710.2, 2262.8, 3536.7, 6163.4, 11087.3, 24043.4 G$_0$ Right panel: Disk fraction vs. FUV flux in the bins of the same grid, showing only the values measured in bins with the relative error of the disk fraction smaller than 0.25}
        \label{dfmapfuvimg}
        \end{figure*}

\begin{figure*}[]
        \centering
        \includegraphics[width=8.0cm]{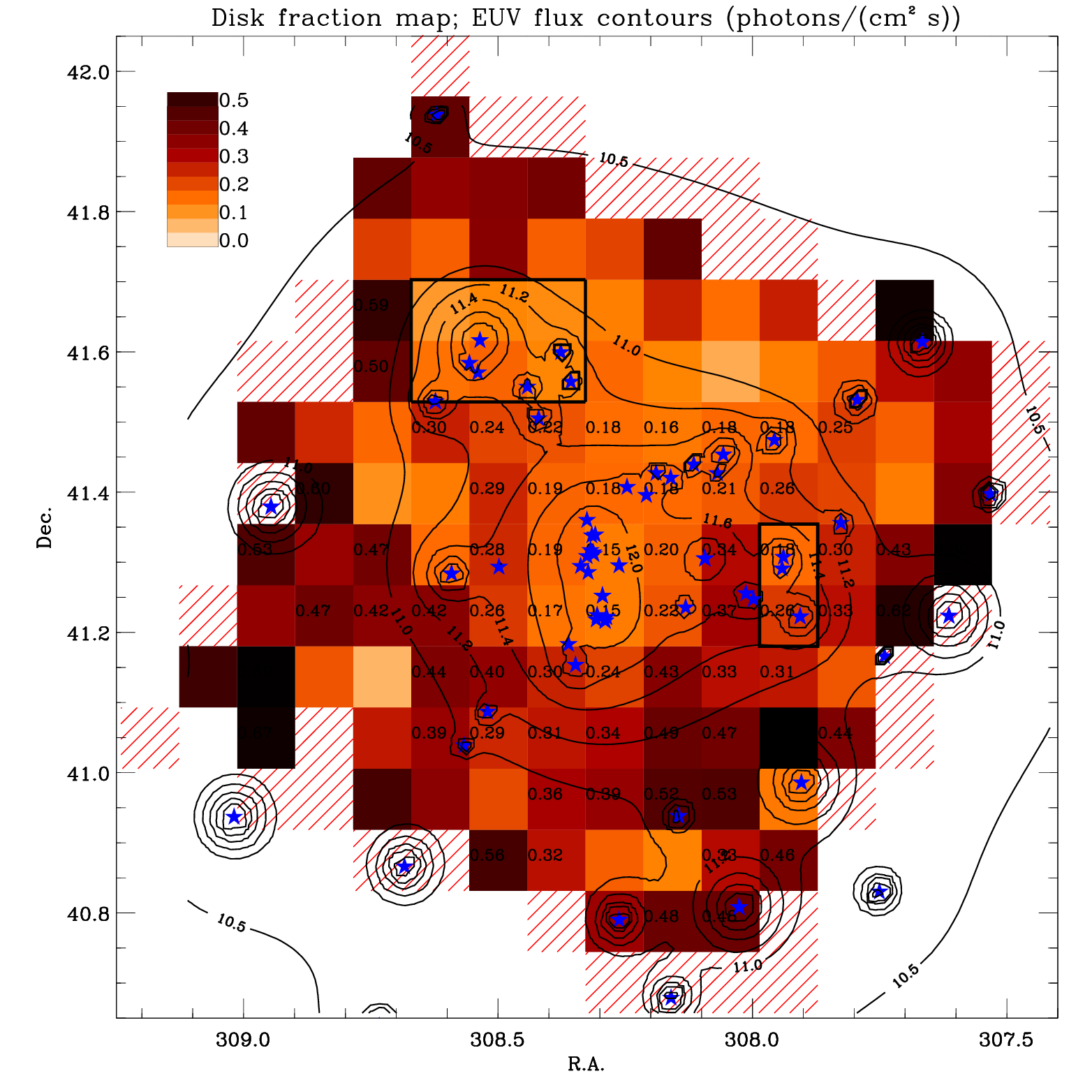}
        \includegraphics[width=8.4cm]{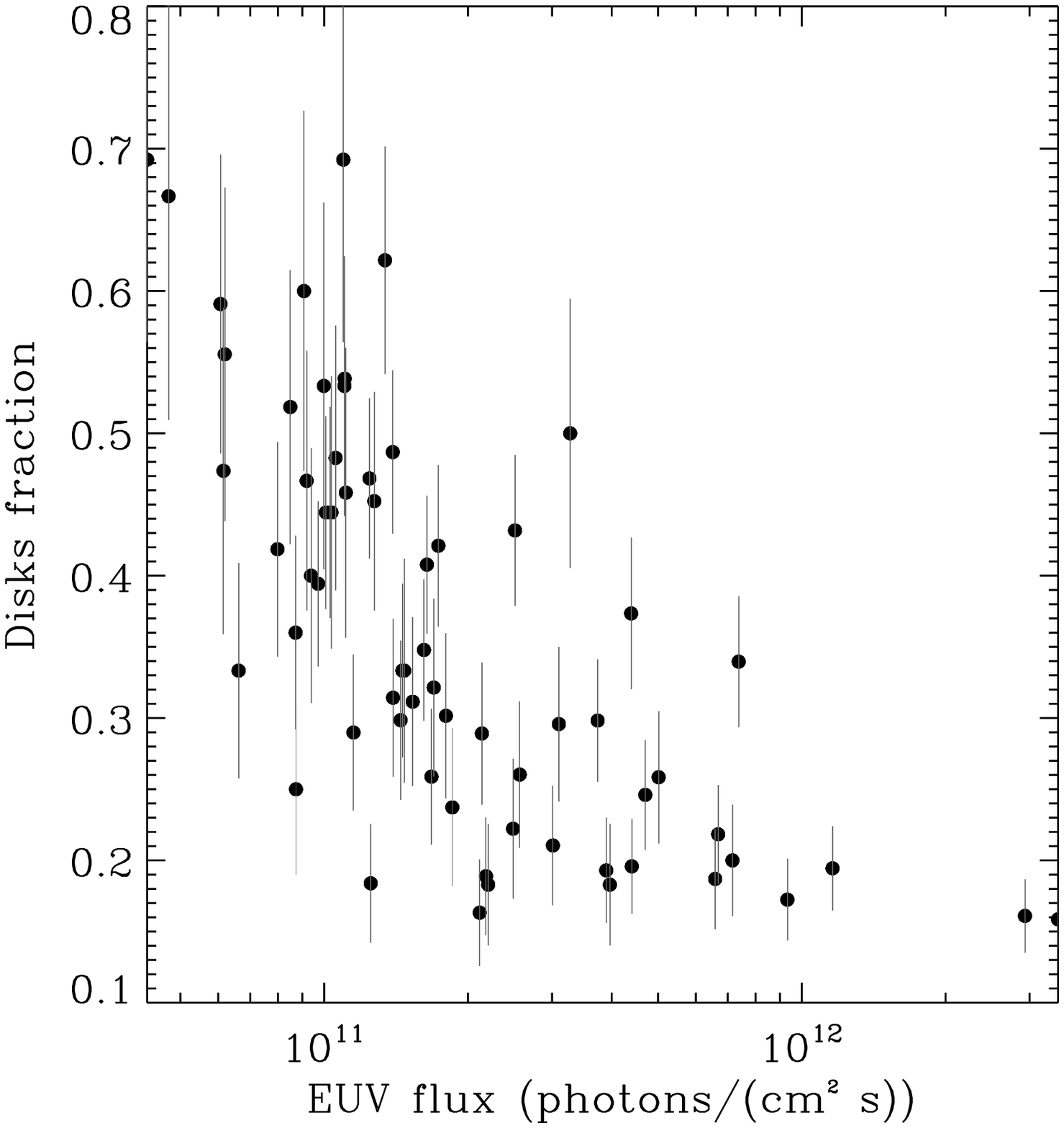}
        \caption{Left panel: Maps showing the spatial variation of the disk fraction as in Fig. \ref{dfmapirimg}. The overplotted contours mark the local EUV flux. The contour levels are: $2.3\times10^{10},\,  6.2\times10^{10},\, 9.3\times10^{10},\, 1.6\times10^{11},\, 3.1\times10^{11},\, 7.7\times10^{11}\,$photons/cm$^{2}$/s. Right panel: Disk fraction vs. EUV flux in the bins of the same grid, showing only the values measured in bins with the relative error of the disk fraction smaller than 0.25}
        \label{dfmapeuvimg}
        \end{figure*}

  In the left panels of Figg. \ref{dfmapfuvimg} and  \ref{dfmapeuvimg} the contours mark the local FUV and EUV fluxes, with the values fixed to the limits of the bins in Fig. \ref{dfuv_img}. By a simple visual inspection of these images, the correlation between the local disk fraction and the local UV field (shown in the right panels of both figures for those bins with low relative error in the disk fraction) is evident. In most of the bins containing one or two O stars, which are characterized by $F_{FUV} > 10^4\,$G$_0$ or log$\left( F_{EUV} \right) > 11.5$ in units of photons/s/cm$^{2}$, the disk fraction is about $15\%\, -\, 20\%$ smaller than that in surrounding bins not containing O stars and characterized by less intense UV fields. Considering that the stellar population of these regions is almost coeval, this gives an estimate of the fraction of disks which have been destroyed by the UV feedback in Cyg~OB2 to date. \par
  
The center of the association, hosting the two central groups of O stars and the elongated group in the north-west direction, is characterized by an almost constant disk fraction ($15\%-18\%$), high surface stellar density ($>33\,$N/pc$^{2}$), and intense UV field ($F_{FUV}>10^4\,$G$_0$ and log$\left( F_{EUV} \right) > 11.5$). This does not allow us to use the data from the central region to compare the efficiency of photoevaporation and collisional destruction. For this aim, the values of the disk fraction in bins hosting isolated O stars are more useful. For instance, in the north-east the region around the group of 7 O stars (marked with a black box in the disk fraction maps) is characterized by intense UV fluxes, but low stellar surface density ranging between $5\,$N/pc$^{2}$ and $17\,$N/pc$^{2}$, and the disk fraction of the whole area is about $15\%$. This is similar to the center of the association in our grid, and significantly larger than the disk fraction observed southward and eastward (also northward, but in this direction the presence of the cloud cavity front and a trunk with ongoing star formation surely affects the disk fraction). Analogously, the disk fractions in the two bins hosting the three O stars at approximately $\alpha\sim307.9$ $\delta\sim41.2$ and $\delta\sim41.3$ (spectral classes O8.5III, O8.5I, and O9.5I), marked with a black box in the disk fraction maps, are 18\% and 26\%.  These values are smaller than those in most of the surrounding bins even if the surface stellar density is similar. \par

\begin{figure}[]
        \centering
        \includegraphics[width=8cm]{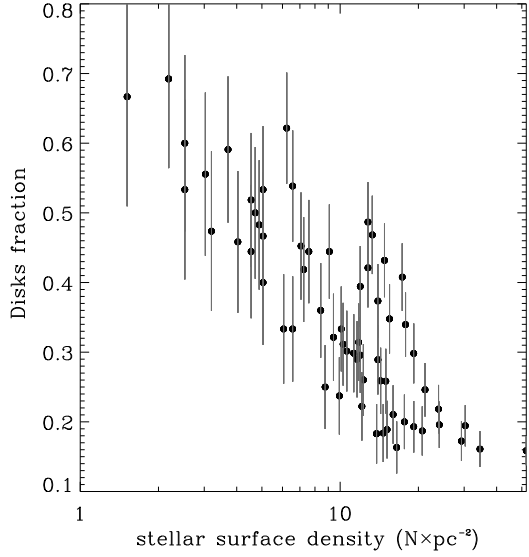}
        \caption{Correlation between the disk fraction and stellar surface density in the bins of the grid used in Fig. \ref{dfmapirimg}.}
        \label{corr_img}
        \end{figure}

The analysis of the disk fraction maps may suggest a greater impact on disk dissipation timescale from the feedback from O stars rather than from close encounters. In fact, this result is only marginally significant and far to be final. However, the spatial variation of disk fraction correlates in a similar way with the local UV field and the stellar surface density, as shown in the Figg. \ref{dfmapdensimg}, \ref{dfmapfuvimg}, and \ref{dfmapeuvimg}, which is not surprising given the strong correlation observed between the local stellar density and the median value of the UV field in these bins (Fig. \ref{corr_img}). Given the difficulty in isolating and comparing regions with different stellar density and similar UV radiation field with regions with similar density and different UV radiation field with good statistic, the analysis described in this section provides only a marginal evidence that disks photoevaporation is a more important mechanism than close encounters in the evolution of disks. \par

	\subsection{Simulated local stellar density}
	\label{density_sec}

In Sect. \ref{2d_sec} we describe the 5000 simulated 3D configurations of the association realized with three different approaches. These simulations allow us to estimate the local stellar density per parsec$^{-3}$ at the position of each member, and to use these estimates to infer the expected rate of stellar encounters. \par

	Given that Cygnus~OB2 is an association by some degree of subclustering, the stellar density in the 3D simulations derived in 3D binning or at increasing radial distances from a nominal center is not useful to characterize the local stellar density experienced by each member. A better estimate can be obtained from the method of \citet{WhitworthBTW1995} and devised for SPH simulations. Considering the $j^{th}$ member of the association whose position is given by its celestial coordinates and the simulated elevations $z$, the local stellar density can be given by $\delta_j=h_j^{-3}\times \sum_{i} W_{ij}$, with $i$ running over the other members, $h_j$ being the smoothing length associated with the $j^{th}$ member, and $W$ an appropriate kernel function. The smoothing length is  defined as the radius of the sphere containing 50 members and centered on the $j^{th}$ member, while $W_{ij}$ is defined as:

 \begin{displaymath}
   W_{ij}(s) = \frac{1}{\pi}\left\{
     \begin{array}{lr}
       1-\frac{3s^2}{2}+\frac{3s^3}{4} & 0\leq s < 1\\
       \frac{\left(2-s \right)^3}{4}   & 1\leq s < 2\\
       0                               & s > 2
     \end{array}
   \right.
\end{displaymath} 

\begin{figure}[]
        \centering
        \includegraphics[width=8cm]{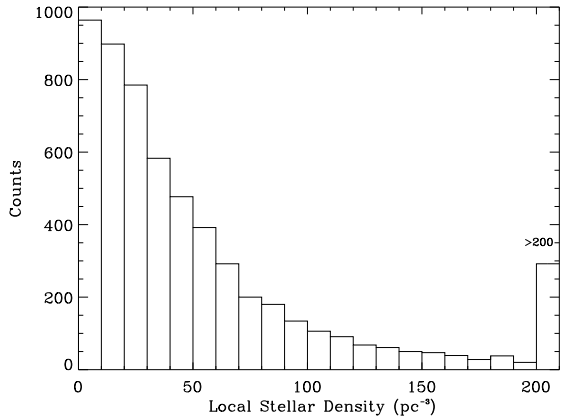}
        \caption{Distributions of local stellar density around the simulated 3D members positions obtained averaging the 5000 simulations described in Sect. \ref{2d_sec}.}
        \label{meddens_img}
        \end{figure}

	{\bf with $s$ being $|r_i-r_j|/h_j$. The average distributions of local stellar densities obtained with the simulations described in Sect. \ref{2d_sec} are shown in Fig. \ref{meddens_img}. The number of members experiencing a local stellar density larger than 200 stars per parsec$^{-3}$ is shown in the last bin. In the simulated 3D stellar configurations there is a significant number (i.e. about 300) of stars with such a large surrounding stellar density. Only for very few members in the simulated configurations do the local stellar densities exceed the values typically observed in open clusters (i.e. from tens to a few hundreds of stars per parsec$^{-3}$).} \par

	The simulated local stellar densities experienced by each association member allow us to make estimates of the chances of close encounters in a given time. For instance, using the calculations presented in \citet{ClarkePringle1991} and adopting a typical disk size of 100$\,$AU, we estimate that densities of $\sim$100 stars per cubic parsec correspond to about 1\% probability of an encounter in 1 Myr. {\bf In the simulated 3D configurations there is a significant fraction of members (14.6\%) surrounded by such a high local stellar density, but even in those cases the 1\% chance of close encounters in 1~Myr is negligible in the context of the disk sample in Cyg~OB2 as a whole.}

{\bf As a further test, we define a fixed grid in the simulated 3D space, with 10 bins per axis, for a total of 1000 cells. For each of the 5000 simulated 3D configurations, we calculate the disk fraction and total stellar density. To infer the total stellar density (i.e. across the entire mass spectrum), we adopt the following procedure. In our X-ray catalog of cluster members, we have 439 stars more massive than 1.5$\,$M$_{\odot}$, 1292 between 1.5$\,$M$_{\odot}$ and 0.7$\,$M$_{\odot}$, and 3098 less massive than 0.7$\,$M$_{\odot}$ \citep{WrightDGA2014,Kashyapinprep}. Using the completeness as a function of stellar mass calculated by \citet{WrightDG2015}, we can predict a total number of members of 488, 1615, and 8551 stars in these three mass intervals, respectively. In order to estimate the number of stars at about 0.1$\,$M$_{\odot}$, we observe that following the IMF defined by \citet{Kroupa01}, the expected population with such a mass is about ten times larger than that with about 0.5$\,$M$_{\odot}$, which in our case means about 85510 stars. Using these numbers, we predict a total population of 96164 stars in Cyg~OB2 down to 0.1$\,$M$_{\odot}$. Incidentally, we note that since the average stellar mass adopting the \citet{Kroupa01} IMF is 0.2$\,$M$_{\odot}$, the total stellar population of Cyg~OB2 we estimate implies a total mass of the association which is very similar to that predicted in existing studies \citep[e.g.][]{WrightDM2015}. To estimate the total stellar density in each cell, we make the assumption that the fraction of members falling in each spatial cell does not depend on stellar mass, i.e. we multiply the fraction of the members falling in each spatial cell by the estimated total number of stars. With this calculation, the estimated stellar density in the 3D grid never exceeds 90 stars per cubic parsec. Such a stellar density is too low to result in significant destructive feedback on disks evolution from the gravitational interaction between members. For instance, \citet{SteinhausenPfalzner2014AA} estimated the number of disks destroyed by close encounters in 2$\,$Myrs in clusters with different stellar density, and no significant effects are observed below stellar densities of 3000 stars per parsec$^3$. Similarly, in the simulations by \citet{VinckeBP2015}, after 5$\,$Myrs in environments with stellar densities of about 90 stars per parsec$^3$ no disks have been destroyed by close encounters below 10$\,$AU, while the fraction of disks affected at radii $\geq100\,$AU from the central star (which is a region of disks that we can not probe with our data) goes from about 10\% to about 17\% with the stellar density increasing from 15 to 90 stars per cubic parsec. In conclusion, the analysis of the simulated 3D stellar distributions of Cyg~OB2 and the comparison between the effects of the local UV field and stellar density indicates that the evolution of protoplanetary disks is impacted by photoevaporation, and that erosion through collisions plays only a minor role.} \par 

\section{Discussion}
\label{disc_sec}

\subsection{Is Cygnus~OB2 a safe environment for disk survival?}
\label{safe_sec}

Externally induced disk photoevaporation has been proven to expedite the dissipation of circumstellar disks. The extraordinary HST optical images of the evaporating proplyds (i.e. protoplanetary disks) in the Trapezium Cluster in Orion \citep{BallySDJ1998} clearly show the impact that photoevaporation induced by nearby O stars (in this case the O6V star $\Theta^1$ Orionis) has on the evolution and morphology of nearby disks. The cometary shape of these evaporating proplyds has been studied by \citet{JohnstoneHB1998} in terms of a neutral flow of gas evaporating from the disk under the influence of the FUV radiation. This gas is then ionized by the EUV radiation forming an ionization front which lies at a distance from the disk surface that depends on several factors, mainly the intensity of the incident UV flux and the optical depth of the evaporating column. The disk mass loss rate of these proplyds has been predicted to range between $\sim10^{-7},\, \sim10^{-8}\,$M$_{\odot}$/yr \citep{StorzerHollenbach1999}, which has been later confirmed by spectroscopic observations and analysis of optical emission lines \citep{HenneyODell1999}. \par
  
  In the Trapezium Cluster photoevaporation is driven by the FUV radiation. When the ionization front lie on the disk surface, however, the EUV radiation can directly ionize the gas in the disk, inducing a more intense mass loss. \citet{StorzerHollenbach1999} have studied the range of intensities of the incident FUV flux produced by the O stars in the Trapezium which results in a photoevaporative flow which is dense enough to absorb the incident EUV photons, i.e. in which the photoevaporation is FUV dominated. With the typical EUV/FUV flux ratio emitted by O stars, they found that the FUV-dominated region occurs for FUV incident fluxes in the range $10^5\,$G$_0 \lesssim F_{FUV} \lesssim 5\times 10^7\,$G$_0$. The lower limit is dictated by the fact that at less intense FUV fluxes the wind is transparent to incident EUV photons that can directly ionize the disk surface, while the upper limit corresponds to distances from the ionizing source with such intense EUV fields that the ionization front coincides with the disk surface. Outside this interval, photoevaporation is EUV-dominated. \par

\begin{figure}[]
        \centering
        \includegraphics[width=9cm]{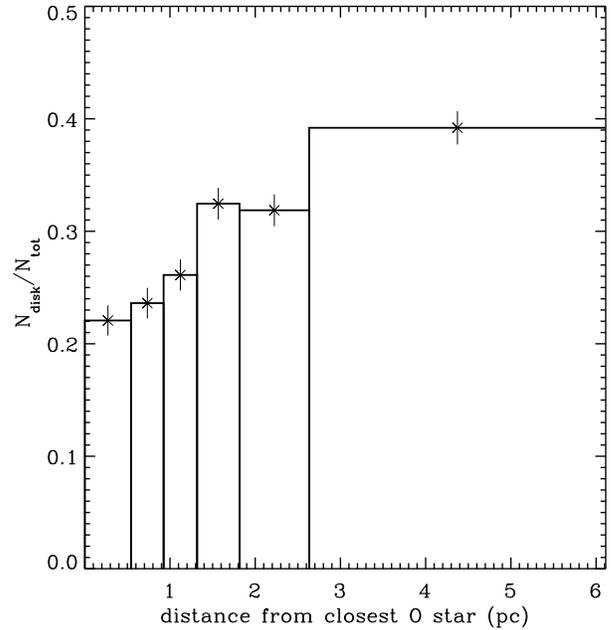}
        \caption{Disk fraction vs. distance of the low mass members from the closest O star.}
        \label{histodist_img}
        \end{figure}

	In Cyg~OB2, the disk fraction in the region irradiated by an FUV flux within the FUV-dominated range is 18.3\%, typical of the very inner area of the association as shown in the disk fraction maps. The distance of these stars from the closest O or WR star (independently from its spectral type) is about 0.4$\,$pc. This can be compared with the decline of the disk fraction with the distance from the closest O star shown in Fig. \ref{histodist_img}. On average, the regions characterized by a disk fraction $\sim20\%$ are less than 1 parsec away from the closest O star. \par

\begin{figure*}[]
        \centering
        \includegraphics[width=9.5cm]{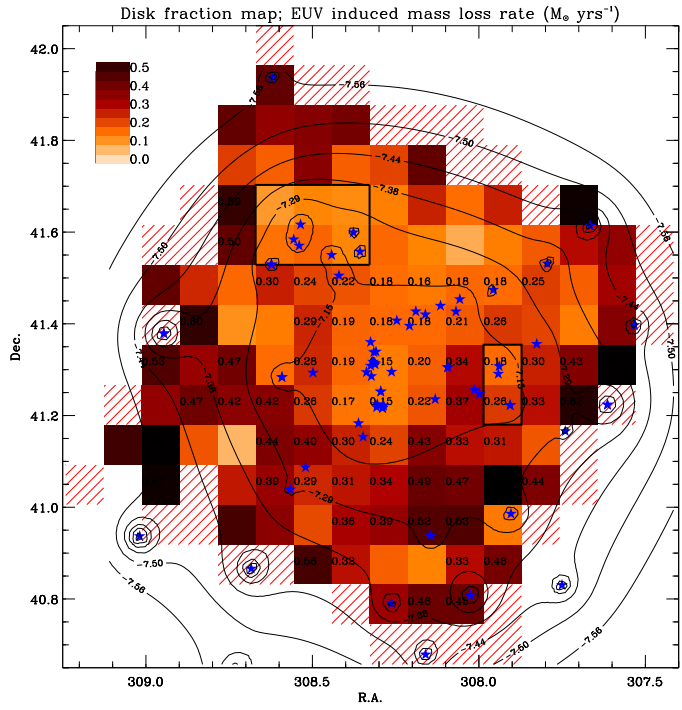}
        \caption{Spatial variation of the disk fraction overplotted with contours marking the expected mass-loss rate due to the photoevaporation externally induced by EUV photons. The label shows the corresponding values of $log \left(\dot{M}\right)$ in units of M$_{\odot}\,$yr$^{-1}$. The disk fraction values are indicated only in those bins with good statistic.}
        \label{mdot_img}
        \end{figure*}

Considering only the photoevaporation induced by FUV photons, this could lead to the conclusion that even in massive associations such as Cygnus~OB2, the externally induced disk photoevaporation is important only nearby the O stars. However, we observe a smooth decline of the disk fraction over the entire region toward the center of the association. To explain this,{\bf we note that a number of simulations \citep[e.g.][]{Clarke2007MNRAS,AndresonAC2013ApJ,FacchiniCB2016MNRAS} suggest that even with incident FUV radiation in the range $3000\,$G$_0 \lesssim F_{FUV} \lesssim 30000\,$G$_0$ the lifetime of protoplanetary disks can be reduced to less than 2$\,$Myrs, unless their viscosity is exceptionally low}. We can also verify whether in regions with $F_{FUV}<10^5\,G_0$ the EUV radiation can drive the photoevaporation of the disks. It is possible to estimate the mass loss rate induced by EUV radiation using the equation derived by \citet{Adams2010}: 
\begin{dmath}
\dot{M}\approx 9\times 10^{-8}\left(\frac{L_{EUV}}{10^{49}\,s^{-1}}\right)^{1/2}\left(\frac{10^{17}\,cm}{d}\right) \times \left(\frac{r_d}{30\,AU}\right)^{3/2}
\label{mdot_eq}
\end{dmath}

where $L_{EUV}$ is the EUV luminosity of the ionizing sources (in photons/s), $d$ is their distance from the disk, and $r_d$ is the disk radius. We calculate the mass-loss rates due to the incident EUV flux emitted by the O stars of Cyg~OB2 adopting $r_d=100\,$AU. The result is shown in Fig. \ref{mdot_img}, where we plot over the disk fraction map the contours of the expected mass loss rate induced by the local EUV radiation field. The contour labels show the log$\left(\dot{M} \right)$ values in units of M$_{\odot}$/yrs. The mass loss rates induced by the extreme O population in Cyg~OB2 ranges from $1.5\times10^{-8}\,$M$_{\odot}$/yr to $3.9\times10^{-7}\,$M$_{\odot}$/yr across the entire field. Such values are capable of dissipating a disk with a typical mass of $0.05\,$M$_{\odot }$ in 3.3 and 0.1 Myrs, respectively. For comparison, a mass loss rate larger than $1.5\times10^{-8}\,$M$_{\odot}$/yr can be induced by $\Theta^1$ Ori only within $0.44\,$pc. This distance is just 0.03 times the projected radius of our survey area. The question that must be addressed now is: why we still observe stars with disks in Cyg~OB2 given this intense mass loss rate induced by the EUV radiation across the entire field? \par

  One possibility is that we are overestimating the incident EUV flux because of the use of the projected distances rather than the real ones. However, in order to dissipate the disks in Cyg~OB2 in a timescale comparable with the association lifetime, the mass loss rate must be about 5 times smaller. This corresponds to an increase in distance $d$ of a factor 5 for the same emitted EUV flux, resulting to unreasonable distances between the outer part of the association and its center. \par
  
Such intense mass loss rates as those we calculate can also be a consequence of the adopted stationary value for $r_d$. Given that the surface density in circumstellar disks decreases with increasing distance from the central star, externally induced photoevaporation, which results in an almost uniform mass loss rate across the disk, dissipates the disk from outside inward \citep{JohnstoneHB1998}. With time, the disk radius shrinks down to the gravitational radius\footnote{The gravitational radius is the maximum distance from the central star where the evaporating gas is gravitationally bound to the system.}. After reducing the disk outer radius to, for instance, 35$\,$AU, the mass loss rate has been reduced by a factor of $\sim$5, following Eq. \ref{mdot_eq}. It must be noted that a such small disk can still produce detectable NIR excesses. \par
  
  Another likely explanation is that Equation \ref{mdot_eq} does not account for the attenuation of the EUV radiation. Two contributions of such absorption are likely to play an important role. First, the calculation of the mass-loss rate in Eq. \ref{mdot_eq} accounts only for the wind produced by the externally induced photoevaporation, ignoring any contribution from the photoevaporation induced by the central star itself (by both X-ray and UV photons). As discussed, this process is thought to be important in the evolution of disks and to be the main process leading to the formation of pre-transition and transition disks \citep[e.g.:][]{ClarkeGS2001,AlexanderCP2004}. However, this contribution is expected to be not dependent on quantities which decrease with the distance from the O stars, such as disks and central stars properties. It is more likely that the presence of a large population of stars with disks in the outer part of the association is instead possible thanks to the absorption of the UV radiation by intracluster material. There is evidence that the inner cluster of Cyg~OB2 is almost clear of gas \citep{SchneiderBSJ2006}. This is not surprising given that massive stars in clusters usually create cavities in the parental cloud in about $3\, -\, 5\,$Myr (e.g. \citealp{AllenMGM2007}). However, H$\alpha$ images of the center of Cyg~OB2 \citep{DrewGIA2005,GuarcelloDWD2013} shows diffuse H$_{\alpha}$ emission around the O stars that must be emitted by intracluster gas. There is also evidence for the presence of diffuse dust emission and dense clumps mainly in the outer part of the association (see Fig. \ref{field_img} and \citealp{GuarcelloDWD2013}). There is, then, a column of absorbing material that partially shields the stars with disks in the outskirt of the association from the UV radiation emitted by the O members, and this extinction is expected to increase with the distance between stars with disks and the O stars. For instance, a decrease of the mass loss rate by a factor of 5 is enough for having disks dissipation timescale comparable with the age of the association. Since, in Eq. \ref{mdot_eq}, $\dot{M}\propto \sqrt{F_{EUV}}$, this requires a decrease by a factor of 25 of the incident UV flux. Dust absorption can be accounted using the extinction law of \citet{CardelliCM1989}: For the EUV radiation ($\lambda<0.125\,\mu m$) a similar absorption can be achieved with an extinction $A_V=1^m$, while in FUV ($\lambda<0.18\,\mu m$) with $A_V=1.4^m$. Using the relation $N_H/A_V=1.8\times 10^{21}\,$atoms/cm$^{2}$/mag, and assuming a constant column density along a distance of 2 or 3 pc, this extinction corresponds to a particle density of $\sim300\,$cm$^{-3}$, which is still typical of giant molecular clouds \citep{SolomonRBY1987}, and it looks realistic given the presence of intracluster material mainly in the outskirt of the association. Gas absorption is even more efficient. We calculate that residual intracluster gas with column density of $\sim10^{19}\,$atoms/cm$^{2}$ is opaque to EUV radiation with $\lambda<912\,$\AA{}: a column density of $1.8\times10^{18}\,$atoms/cm$^{2}$ corresponds to a transmittance (the fraction of incident flux which is not absorbed by the gas) of $\sim0.004$, becoming about two orders of magnitude smaller by increasing the gas column density of a factor of two. Considering that the hydrogen (H+H$_2$) column density that we can calculate from Herschel data is larger ($\sim10^{21}\,$atoms/cm$^{2}$), even if part of this comes from the Rift in the foreground, the  absorption of the EUV radiation by residual intracluster material is then the most reliable hypothesis to explain the presence of a larger fraction of members with disk in the outer part of the association with respect to the center. \par

\subsection{Cygnus~OB2 in context}
\label{comp_sec}

In the previous sections, we have found evidence that in Cygnus~OB2 disk evolution has been seriously affected by the surrounding environment. \par

\begin{figure}[]
        \centering
        \includegraphics[width=8.5cm]{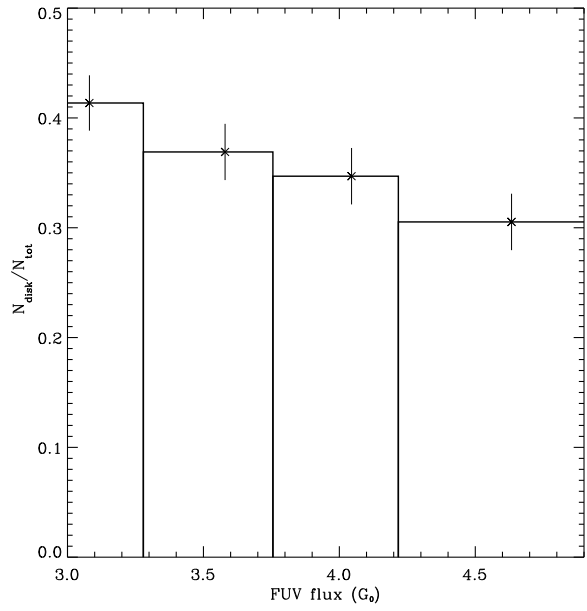}
        \caption{Disk fraction vs. FUV flux as observed in NGC~6611.}
        \label{6611_img}
        \end{figure}

  Similar effects have been studied in other clusters with different properties. In their series of papers, \citet{GuarcelloPMD2007, GuarcelloMDP2009, GuarcelloMPP2010} have found that the spatial variation of disk fraction in NGC~6611 is constant across the cluster and equal to $\sim40\%$, except for a sudden decrease down to $26\%$ within $1\,$pc from the massive stars (O plus early B). NGC~6611 is an intermediately massive cluster hosting 54 OB stars, among which are 13 O stars mainly concentrated in a cavity $2.2\,$pc in diameter \citep{HillenbrandMSM1993}. The most massive star in this cluster is W205 with a mass of $75-80\, M_{\odot}$ and a spectral class O3-O5V \citep{EvansSLL2005}. NGC~6611 is younger than Cyg~OB2, having a median age of $\sim1\,$Myr \citep{GuarcelloPMD2007}. In Fig. \ref{6611_img} we recalculate the variation of the disk fraction in NGC~6611 as a function of the incident FUV flux (in terms of G$_0$) as we did in this paper for Cygnus~OB2. The difference between the disk fraction in the first and fourth bins is the only significant difference we observe, with the disk fraction going from $\sim41\%$ to $\sim31\%$. Compared to Cyg~OB2, the disk fractions in the outer population of the two regions are similar, while the main difference is observed in the central parts, close to the O stars. The differences between the two regions are likely a consequence of the different ages of the two stellar populations, with a combination of normal disk evolution and induced photoevaporation\footnote{The hypothesis of an inside-out star formation chronology in NGC~6611 has been discarded by \citet{GuarcelloMPP2010}}. However, in the recent paper of \citet{RichertFGK2015} no evidence of rapid disk dissipation nearby the massive stars in NGC~6611 is found. These authors claim that the different result is due to the different membership selection, since their selection of stars with excesses in the NIR bands is based on UKIDSS data while that in \citet{GuarcelloMDP2009} used 2MASS data, allowing them to observe less massive stars and disks. Since the decrease of the disk fraction in the center of NGC~6611 was observed also by \citet{GuarcelloMPP2010}, in which stars with excesses in $JHK$ bands were also selected using UKIDSS data, it is very likely that the disparate results are due to the different approaches both in selecting cluster members and in the analysis of the spatial variation of disk fraction. This simply suggests that the external feedback in disk evolution in clusters such as NGC~6611 is still an open question which requires further analysis. \par  
  
  A similar situation has been found in the coeval cluster Pismis~24 by \citet{FangBKH2012}. This $\sim1\,$Myr old cluster hosts dozens of OB stars, with some very massive stars such as Pis24-1 (O3I) and Pis24-17 (O3.5III) \citep{MasseyDW2001}, together with a rich low mass population. \citet{FangBKH2012} found that in this cluster the disk fraction is constant at $\sim36\%$ across the field, decreasing down to $19\%$ at a distance of $0.5\,$pc from the most massive members. The stars within 0.5$\,$pc from these two massive stars are irradiated by a FUV flux more intense than 21000$\,$G$_0$, which is similar to the FUV flux incident on the closest stars to the O stars in Cyg~OB2 and NGC~6611. The scenario resulting from these studies is similar: in intermediately massive clusters after $1\,$Myr the environment feedback on disk evolution is important only in the immediate proximity of the O and B stars, and generally in the cluster core, while on this timescale no effects are experienced by the disks in the outer region. \par
  
Even for slightly older intermediately massive clusters the situation is similar, as demonstrated by the spatial variation of disk fraction in NGC~2244. This cluster is $2\, -\, 3\,$Myrs old \citep{ParkSung2002} and it hosts 7 O stars and a conspicuous low mass population. In NGC~2244 the average disk fraction is $\sim40\%$, and constant across the cluster but with a rapid decrease down to $\sim23\%$ at distances from the O stars $<0.5\,$pc \citep{BalogMRS2007}. \par

	Another important contribute in this context are the $N$-body simulations of the evolution of protoplanetary disks in the Orion Nebula Cluster by \citet{ScallyClarke2001}. They followed the evolution of the disks with radii of 100$\,$AU and 10$\,$AU for $10^7\,$years in a cluster with 4000 members, considering the feedback provided by both the UV radiation and stellar encounters separately. They have found that photoevaporation removes between 0.01$\,$M$_{\odot}$ and 1$\,$M$_{\odot}$ from the larger (100$\,$AU) disks, while the 10$\,$AU disks are not seriously affected. Similarly, they estimated that only 4\% of the stars have encounters closer than 1000$\,$AU. These studies confirms then that important effects on disks evolution in such environments can be experienced only in the dense cluster core, characterized by intense UV field, even for timescales larger than $1\,$Myr. \par
  
  In clusters with small populations of massive stars and low stellar density no important feedback on the evolution of circumstellar disks is expected. The local UV field can reach critical intensities only at very small distances from the few massive members.  Besides, as noted in Sect. \ref{intro}, in clusters hosting a few hundred members the chances of close encounters with small ($\sim$100$\, -\, 200\,$AU) impact parameters is $\leq1\%$ in $1\,$Myr. This is confirmed by studies of the spatial variation of the disk fraction in IC~1795 \citep{RoccatagliataBHG2011}. This $\sim3\,$Myr old cluster hosts two O stars (O6.5V and O9.7I), and no variation of disk fraction has been observed toward the position of these ionizing sources. This can be understood in terms of the FUV flux produced by these sources. For instance, the median FUV flux experienced by the stars in Cyg~OB2 lying in the first bin in Fig. \ref{dfuv_img} (i.e. the one with the highest UV fluxes, containing more than 700 candidate members) is $\sim54100\,$G$_0$. The two O stars in IC~1795 only produce such intense FUV flux within 0.2$\,$pc, where 9/525 candidate members lie. In their search for accreting objects in the young cluster IC~1396, around the O6.5V star HD~206267, \citet{BarentsenVDG2011} found that mass accretion rates, number of accretors, and intensity of the infrared excesses increase marginally with the distance from this massive star, interpreting this result as a consequence of triggered star formation rather than induced photoevaporation. \par

All these results taken together indicate that most of the environments where stars formation occurs in our Galaxy are safe for disk evolution and planet formation: in sparse clouds and low mass clusters, as well as in the outer regions of intermediately massive clusters, the local FUV and EUV fluxes and the local stellar densities never exceed values that would cause a rapid erosion of circumstellar disks, potentially halting or disrupting the planet formation process. Important feedback is expected and observed only in the core of intermediately massive clusters, at distances of $\ll 1\,$pc from the massive stars that are usually found in the cluster center. \par

  There is strong evidence showing that even our Sun and Solar System formed in the outer part of an intermediately massive cluster. \citet{Adams2010} noted that the orbits of the planets in the Solar System require no close encounters with $b\leq90\,$AU, indicating that the parental cluster of the Sun had less than $10^4$ members; at the same time, the orbits of some trans-neptunian objects such as Sedna require at least one encounter with $200\,$AU$\leq b \leq 300\,$AU; the paucity of gas in the trans-Neptunian objects and the presence of giant planets require the incident FUV flux to have been in the range $2000\,$G$_0 \leq F_{FUV} \leq 10^4\,$G$_0$; the presence in the Solar System today of short-lived radio nuclei with half-life $<$ few Myrs (such as $^{26}$Al and $^{30}$Fe) requires the presence of O stars. All these factors indicate that the parental cluster of the Sun was an intermediately massive cluster hosting a few thousand members, and that the Sun formed in its periphery, and eventually moved toward its center \citep{Adams2010}. The outer regions of clusters like Pismis~24 and NGC~6611 are good examples of such star forming regions, which are quite common in our Galaxy. \par

In this paper we complete this picture by studying the feedback provided in very massive associations that, even if rare in our Galaxy, harbor tens of thousands of stars. Our study indicates that in environments such as Cygnus~OB2 the local values of the EUV and FUV radiation field are intense enough to externally induce the photoevaporation of disks even at large distances from the O stars ($\leq 10\,$pc), and likely exceeding the influence of close encounters, given the moderate stellar density observed across the region. Using the estimate from existing and detailed models of the photoevaporation process induced by nearby O stars, photoevaporation is driven by the FUV radiation in regions characterized by FUV fluxes $\ge 10^5\,$G$_0$, i.e. at distances $\leq 0.5\,$pc from the O stars, and then by the EUV photons in the rest of the association. The induced mass loss rate from the EUV flux produced by the massive population in Cyg~OB2 would be intense enough to completely dissipate the disks across the entire association very quickly. The presence to date of a large disks population is likely a consequence of the column of extinguishing material between the low-mass and the O members of Cyg~OB2, that absorbs the EUV radiation and then shields the disks. This likely plays also a role in shaping the observed smooth decline of the disk fraction with the distance from the center of the association. Our data suggest that the overall effect on the timescale corresponding to the age of Cyg~OB2 members (i.e. between $3\,$Myrs and $5\,$Myrs) of the environmental feedback on disks evolution is the decrease of the disk fraction by about $20\%$. \par
  
  While disk erosion is likely to play a role in the planet formation process, the effect it has is not yet known. The observation of photoevaporating disks at $8.0\mu m$ in NGC~2244 \citep{BalogRSM2006} indicates that small dust grains remain trapped in the photoevaporating flow, reducing the reservoir of solid material available for the formation of planetesimals in the disk. On the other hand \citet{OwenEC2011} show that only a small fraction of the existing small dust remain trapped in the photoevaporative flow, and the simulations presented by \citet{ThroopBally2005} indicate that in the midplane of gas-depleted disks gravitational instabilities occur more easily, allowing a rapid formation of planetesimals with dimensions from centimeters to kilometers. More work is necessary to better assess the true impact of rapid photoevaporation on planet formation. \par
  
\section{Summary and conclusions}
\label{conclusions}

In this paper we study how the environment in Cygnus~OB2 affects the dissipation timescale of protoplanetary disks, by analyzing the spatial variation of the disk fraction across the association. We use the selection of members with and without disks provided by other publications related to the $Chandra$ Cygnus~OB2 Legacy Survey. We correlate the local values of disk fraction across the association with the local values of EUV and FUV fluxes, and stellar surface density, observing a smooth decline of the disk fraction from $\sim40\%$ to $\sim18\%$ with increasing UV fluxes and stellar density. \par
  
  We rule out the hypothesis that the observed decline of disk fraction across the association is a consequence of an inside-out triggering of star formation, a 2D projection effect, or non uniform sensitivity of our data. We also briefly discuss the existing evidence supporting the hypothesis that the association is not dynamically evolved and never been much denser than we observe today.  \par
  
We interpret the result as a consequence of the rapid erosion of the disks in Cyg~OB2 by the incident UV radiation. In particular, using models of externally induced photoevaporation, we conclude that FUV radiation dominates the process in the region within $\sim 0.5\,$pc from the O stars. However, the EUV field is intense enough to induce the dissipation of the disks in few Myrs across the entire association. The presence of a significant fraction of stars with disks in the outer regions and the observed smooth decline of the disk fraction with the intensity of the incident UV flux is explained in terms of an absorption of the EUV radiation by the material still associated with the cloud whose efficiency increases at increasing distances from the O stars. We find the destruction of disks by close stellar encounters to be rare, such that only of the order of 1\% or fewer disks in the regions of the association characterized by a stellar density larger than 100 stars/pc$^3$ per Myr. \par

  Finally, we consider similar studies published so far on clusters with different size and age. Only the core of intermediately massive clusters are characterized by UV fluxes and stellar densities that have a strong impact on disks evolution, while the outer regions are relatively safe environments for disk evolution. Analogously, massive associations are harsh environments only if intracluster material can not provide efficient shielding from the ionizing radiation emitted by the massive stars. These results suggest that a large variety of star forming environments in our Galaxy are safe environments for disk evolution and planet formation (i.e.: sparse clouds, low mass clusters and the outer regions of intermediately massive clusters).


\acknowledgments
We thank the referee for his/hers suggestions and help improving our manuscript. MGG acknowledges the grant PRIN-INAF 2012 (P.I. E. Flaccomio). NJW acknowledges a Royal Astronomical Society Research Fellowship. They also have been supported by the Chandra grant GO0-11040X during the course of this work. JJD and VK were supported by NASA contract NAS8-03060 to the Chandra X-ray Center, and thank the Director, B. Wilkes, and the CXC science team for advice and support.
%
\addcontentsline{toc}{section}{\bf Bibliografia}
\bibliographystyle{apj}
\bibliography{biblio}


 \onecolumn
 \begin{center}
\begin{longtable}{p{2.5cm}p{2.5cm}p{2.5cm}p{2.5cm}p{2.5cm}}
\caption[]{UV fluxes emitted by the O stars in the Cyg~OB2.} \label{ostars_tb} \\

\hline \hline \\ RA & DEC & Spectral Type & log$\left( L_{FUV}/L_{\odot} \right)$ & log$\left( L_{EUV} \right)$ \\
       J2000 & J2000& $ $ & erg/s & photons/s \\ 
       \hline \vspace{0.05cm}
\endfirsthead

\multicolumn{5}{c}{{\tablename} \thetable{} -- Continued} \\
  \hline \hline \\RA & DEC & Spectral Type & log$\left( L_{FUV}/L_{\odot} \right)$ & log$\left( L_{EUV} \right)$ \\
       J2000 & J2000& $ $ & erg/s & photons/s \\
       \hline \vspace{0.05cm}
\endhead

  \multicolumn{5}{l}{{Continued on Next Page\ldots}} \\
\endfoot

  \\ \hline \hline 
\endlastfoot

20:33:08.779   &   +41:13:18.10    &  O3If    &  5.38  &  49.79 \\	  
20:33:08.779   &   +41:13:18.10    &  O6V     &  5.06  &  48.96 \\	  
20:33:14.160   &   +41:20:21.50    &  O3I     &  5.33  &  49.79 \\	  
20:33:18.019   &   +41:18:31.00    &  O5III   &  5.25  &  49.44 \\	  
20:33:10.740   &   +41:15:08.00    &  O5I     &  5.40  &  49.60  \\	  
20:33:10.740   &   +41:15:08.00    &  O3.5III &  5.42  &  49.64 \\	  
20:34:08.539   &   +41:36:59.30    &  O5I     &  5.28  &  49.61 \\	  
20:33:23.460   &   +41:09:12.90    &  O5.5V   &  5.40  &  49.11 \\	  
20:33:13.250   &   +41:13:28.60    &  O6V     &  4.96  &  48.96 \\	  
20:31:37.500   &   +41:13:21.10    &  O6IV    &  5.19  &  49.09 \\	  
20:31:37.500   &   +41:13:21.10    &  O9III   &  4.75  &  48.42 \\	  
20:33:15.180   &   +41:18:50.10    &  O5.5III &  5.21  &  49.33 \\	  
20:33:15.180   &   +41:18:50.10    &  O6I     &  5.24  &  49.44 \\	  
20:33:14.839   &   +41:18:41.40    &  O6.5III &  5.14  &  49.11 \\	  
20:34:44.100   &   +40:51:58.00    &  O6.5III &  5.00  &  49.11 \\	  
20:33:40.879   &   +41:30:18.50    &  O7V     &  4.71  &  48.63 \\	  
20:34:13.500   &   +41:35:02.60    &  O7V     &  4.84  &  48.63 \\	  
20:34:29.520   &   +41:31:45.50    &  O7V     &  4.98  &  48.63 \\	  
20:34:29.520   &   +41:31:45.50    &  O9V     &  4.49  &  47.90 \\	  
20:31:59.611   &   +41:14:50.40    &  O7V     &  4.76  &  48.63 \\	  
20:32:13.771   &   +41:27:12.70    &  O7III   &  4.96  &  49.00 \\	  
20:32:22.430   &   +41:18:19.00    &  O7I     &  5.24  &  49.27 \\	  
20:32:22.430   &   +41:18:19.00    &  O6I     &  5.32  &  49.44 \\	  
20:32:22.430   &   +41:18:19.00    &  O9V     &  4.49  &  47.90 \\	  
20:31:36.910   &   +40:59:09.10    &  O7I     &  5.07  &  49.27 \\	  
20:30:27.300   &   +41:13:25.00    &  O7I     &  4.82  &  49.27 \\	  
20:30:27.300   &   +41:13:25.00    &  O9I     &  5.07  &  48.80 \\	  
20:33:17.489   &   +41:17:09.20    &  O7.5V   &  4.84  &  48.44 \\	  
20:33:26.770   &   +41:10:59.50    &  O7.5V   &  4.76  &  48.44 \\	  
20:32:31.500   &   +41:14:08.00    &  O7.5III &  5.13  &  48.89 \\	  
20:31:10.570   &   +41:31:53.00    &  O8V     &  4.85  &  48.29 \\	  
20:32:27.670   &   +41:26:21.70    &  O8V     &  4.69  &  48.29 \\	  
20:32:38.580   &   +41:25:13.60    &  O8V     &  4.76  &  48.29 \\	  
20:32:45.451   &   +41:25:37.30    &  O8V     &  4.82  &  48.29 \\	  
20:32:59.170   &   +41:24:25.70    &  O8V     &  4.65  &  48.29 \\	  
20:33:02.940   &   +41:17:43.30    &  O8V     &  4.76  &  48.29 \\	  
20:33:13.670   &   +41:13:05.70    &  O8V     &  4.67  &  48.29 \\	  
20:33:18.079   &   +41:21:36.60    &  O8V     &  4.69  &  48.29 \\	  
20:33:30.430   &   +41:35:57.50    &  O8V     &  4.82  &  48.29 \\	  
20:32:34.800   &   +40:56:17.00    &  O8V     &  4.62  &  48.29 \\	  
20:34:21.950   &   +41:17:01.60    &  O8III   &  4.58  &  48.76 \\	  
20:34:21.950   &   +41:17:01.60    &  O8III   &  4.58  &  48.76 \\	  
20:33:02.899   &   +40:47:25.00    &  O8II    &  5.13  &  48.90 \\	  
20:32:50.030   &   +41:23:44.60    &  O8.5V   &  4.66  &  48.10 \\	  
20:33:16.361   &   +41:19:01.90    &  O8.5V   &  4.61  &  48.10 \\	  
20:33:21.041   &   +41:17:40.10    &  O8.5V   &  4.53  &  48.10 \\	  
20:33:25.670   &   +41:33:26.60    &  O8.5V   &  4.76  &  48.10 \\	  
20:31:45.389   &   +41:18:26.80    &  O8.5I   &  4.77  &  48.95 \\	  
20:31:18.310   &   +41:21:21.70    &  O9V     &  4.52  &  47.90 \\	  
20:32:16.531   &   +41:25:36.40    &  O9V     &  4.55  &  47.90 \\	  
20:33:09.581   &   +41:13:00.60    &  O9V     &  4.39  &  47.90 \\	  
20:34:04.949   &   +41:05:13.20    &  O9V     &  4.46  &  47.90 \\	  
20:34:09.521   &   +41:34:13.40    &  O9V     &  4.49  &  47.90 \\	  
20:33:09.410   &   +41:12:58.20    &  O9V     &  4.32  &  47.90 \\	  
20:31:49.651   &   +41:28:26.80    &  O9III   &  4.42  &  48.42 \\	  
20:33:46.150   &   +41:33:00.50    &  O9I     &  5.18  &  48.80 \\	  
20:33:15.739   &   +41:20:17.20    &  O9.5V   &  4.42  &  47.56 \\	  
20:33:59.570   &   +41:17:36.10    &  O9.5V   &  4.45  &  47.56 \\	  
20:30:57.701   &   +41:09:57.00    &  O9.5V   &  4.53  &  47.56 \\	  
20:34:16.049   &   +41:02:19.60    &  O9.5V   &  4.47  &  47.56 \\	  
20:31:46.001   &   +41:17:27.40    &  O9.5I   &  4.41  &  48.69 \\	  
20:30:39.869   &   +41:36:50.90    &  O6V     &  5.06  &  48.96 \\	  
20:34:55.099   &   +40:34:43.97    &  O9V     &  4.49  &  47.90 \\	  
20:30:07.800   &   +41:23:49.99    &  O8V     &  4.69  &  48.29 \\	  
20:32:38.400   &   +40:40:43.97    &  O8III   &  4.90  &  48.76 \\	  
20:34:55.999   &   +40:38:17.99    &  O9.7I   &  5.04  &  48.69 \\	  
20:32:30.300   &   +40:34:32.99    &  O9.5IV  &  4.53  &  47.93 \\	  
20:36:04.500   &   +40:56:12.98    &  O5V     &  5.18  &  49.26 \\	  
20:31:00.101   &   +40:49:48.97    &  O7V     &  4.89  &  48.63 \\	  
20:27:37.870   &   +41:15:46.80    &  O9.5V   &  4.40  &  47.56 \\	  
20:34:28.999   &   +41:56:16.98    &  O9V     &  4.49  &  47.90 \\	  
20:32:03.101   &   +41:15:19.90    &  WC4     &  3.77  &  49.10 \\	  
20:32:06.281   &   +40:48:29.70    &  WN7     &  4.64  &  49.40 \\	  
20:32:06.281   &   +40:48:29.70    &  O7V     &  4.92  &  48.63 \\	  
20:35:47.090   &   +41:22:44.70    &  WC6     &  4.23  &  49.10 \\	  
20:35:47.090   &   +41:22:44.70    &  O8III   &  5.10  &  48.76 \\	  

\end{longtable}
\end{center}

%
%
%
%

 \clearpage
 \twocolumn

         \clearpage

         \clearpage

         \clearpage

         \clearpage

         \clearpage

         \clearpage

         \clearpage

        \clearpage

          \clearpage

          \clearpage

          \clearpage
        

\end{document}